  \documentclass[twocolumn]{aastex63}

 \usepackage{longtable}

\newcommand{\Htwo}{H$_2$}

\newcommand {\HI}     {\ion{H}{1}}      
\newcommand {\HII}    {\ion{H}{2}}      
\newcommand {\HeII}   {\ion{He}{2}}   

\newcommand {\CI}       {\ion{C}{1}}           
\newcommand {\CIII}     {\ion{C}{3}}           
\newcommand {\CIV}    {\ion{C}{4}}           

\newcommand {\NIII}      {\ion{N}{3}}
\newcommand {\NIV}      {\ion{N}{4}}

\newcommand {\OVI}     {\ion{O}{6}}      

\newcommand {\MgII}    {\ion{Mg}{2}}

\newcommand {\SiII}      {\ion{Si}{2}}
\newcommand {\SiIII}     {\ion{Si}{3}}
\newcommand {\SiIV}    {\ion{Si}{4}}

\newcommand {\SII}      {\ion{S}{2}}
\newcommand {\FeII}    {\ion{Fe}{2}}
\newcommand {\ZnII}    {\ion{Zn}{2}}

\newcommand {\AlIII}    {\ion{Al}{3}}

\newcommand{\IUE} {{\it IUE}} 
\newcommand{\FUSE}{{\it FUSE}}
\newcommand{\HST} {{\it HST}} 

\newcommand {\Lya}    {Ly$\alpha$}   

\newcommand\etal{et~al.}

\begin{document}

\title{\large Distances to Galactic OB-stars:  Photometry vs. Parallax } 

\author{ J. Michael Shull and Charles W. Danforth}
\affiliation{CASA, Dept.\ of Astrophysical and Planetary Sciences  \\
University of Colorado, 389-UCB, Boulder, CO 80309}

\email{michael.shull@colorado.edu, danforth@colorado.edu}


\begin{abstract}

For application to surveys of  interstellar matter and Galactic structure, we compute new 
spectrophotometric distances to 139 OB stars frequently used as background targets for UV 
spectroscopy.  Many of these stars have updated spectral types and digital photometry with 
reddening corrections from the Galactic O-Star (GOS) spectroscopic survey.  We compare
our new photometric distances to values used in previous \IUE\ and \FUSE\ surveys and to 
parallax distances derived from {\it Gaia}-DR2, after applying a standard (0.03 mas) offset from 
the quasar celestial reference frame.   We find substantial differences between photometric and 
parallax distances at $d > 1.5$~kpc, with increasing dispersion when parallax errors exceed 8\%.  
Differences from previous surveys arise from new GOS stellar classifications, especially luminosity 
classes, and from reddening corrections.  We apply our methods to two OB associations.   
For Perseus OB1 (nine O-stars) we find mean distances of $2.47\pm 0.57$~kpc ({\it Gaia} parallax) 
and $2.99 \pm 0.14$~kpc (photometric) using a standard grid of absolute magnitudes (Bowen \etal\ 
2008).  For 29 O-stars in Car~OB1 associated with Trumpler-16, Trumpler-14, Trumpler-15, and
Collinder-228 star clusters, we find $2.87 \pm 0.73$~kpc ({\it Gaia}) and $2.60 \pm 0.28$~kpc
(photometric).  Using an alternative grid of O-star absolute magnitudes (Martins \etal\ 2005) shifts
these photometric distances $\sim$7\% closer.  Improving the distances to OB-stars will require
attention to spectral types, photometry, reddening, binarity, and the grid of absolute magnitudes.  
We anticipate that future measurements in {\it Gaia}-DR3 will improve the precision of distances 
to massive star-forming regions in the Milky Way.

\vspace{1cm}

\end{abstract}


\section{INTRODUCTION}  

Quantitative analyses of the structure of the Milky Way galaxy (Binney \& Merrifield 1998) and its interstellar 
medium (ISM) depend on knowing the distances to stars that map out their positions and motions.  These 
distance estimates began with parallax measurements for local stars and were later extended to photometric 
distance estimates.  Space-astrometric missions ({\it Hipparcos}, {\it Gaia}) have expanded the horizon for
parallax measurements to stars at kiloparsec scales.  Indeed, many astronomers hoped that {\it Gaia} Data
Release 2 (DR2) would provide accurate distances to large numbers of massive OB-type stars throughout 
the Milky Way (Chan \etal\ 2019).  Similar hopes arose from the Galactic O-Star (GOS) spectroscopic survey 
(Ma\'iz Apell\'aniz \etal\ 2004) which generated a large sample of O stars with updated spectral types 
(Sota \etal\ 2011, 2014) within several kpc of the Sun.  In fact, a systematic discrepancy has appeared between 
photometric and parallax distances, as discussed below. The GOS digital photometry and optical-NIR dust extinction 
(Ma\'iz-Apell\'aniz \& Barb\'a 2018) offer an opportunity to compute new ``spectrophotometric distances" using 
current grids of absolute magnitudes for O-stars (Bowen \etal\ 2008;  Martins \etal\ 2005) which can be 
compared to distances from {\it Gaia}.  

In this paper, we compute new photometric distances to a sample of OB-type stars frequently used in UV 
absorption-line studies of Galactic interstellar gas (\HI, \Htwo, \OVI).  We focus on 139 OB-type stars used
as background targets in our forthcoming \FUSE\ survey of interstellar \Htwo\ absorption.  The goals
of this paper are threefold.  First, we calculate new photometric distances ($D_{\rm phot}$) for these OB stars.  
Second, we estimate parallax distances ($D_{\rm Gaia}$) from {\it Gaia}-DR2, applying a single (0.03 mas)
parallax offset from the celestial reference frame of quasars.  Third, we compare our photometric distances with 
previous estimates and with {\it Gaia} distances.  We evaluate the differences between the methods, including a
critical literature review of spectral types, luminosity classes, photometry, and reddening of all 139 OB stars.   
Our new photometric distances are compared to those in previous UV surveys and to {\it Gaia}-DR2, identifying 
outliers in plots of $D_{\rm phot}$ vs.\ $D_{\rm Gaia}$.  Discrepancies in the ratio, $D_{\rm Gaia} / D_{\rm phot}$,
may arise from dispersion in the parallax offsets or from incorrect SpTs (and absolute magnitudes).   Finally, 
we apply these methods to two Galactic OB associations, Per OB1 and Car OB1.  We investigate whether 
OB-star photometric distances can be used to define cluster membership and characterize the range of 
parallax offsets.

Recent analyses of {\it Gaia}-DR2 (Lindegren \etal\ 2018;  Arenou \etal\ 2018;  Brown \etal\ 2018) found
systematic fluctuations in parallaxes relative to the reference frame of distant quasars.  For example, parallax 
offsets were seen in {\it Gaia} data toward Cepheids (Riess \etal\ 2018), eclipsing binaries (Stassun \& Torres 
2018; Graczyk \etal\ 2019), red-giant stars in the Kepler field with asteroseismic distances (Zinn \etal\ 2019), 
and OB stars in the Carina OB1 Association near $\eta$ Carinae (Davidson \etal\ 2018).  These offsets are 
greater toward bright stars ($G_{\rm Gaia}  < 12$), and they appear to depend on stellar color and location on 
the sky.  We adopt a standard zero-point parallax offset, $\varpi_{\rm ZP} =  0.03$~mas, which we {\it add} to the 
tabulated {\it Gaia} parallax angle ($\varpi$).   Although some studies have found larger offsets (0.05-0.08 mas) 
for redder stellar populations, we choose 0.03 mas as the appropriate color match between OB stars and the 
blue spectra of quasars.  A mean offset of 0.03 mas has also been found in a recent study of eclipsing 
binaries (Graczyk \etal\ 2019).  Because the characterization of parallax offsets remains uncertain, we avoid 
using more complex statistical corrections (Bailer-Jones \etal\ 2018).  Instead, for each star, we apply a 
correction to the parallax angle to find a distance $D_{\rm Gaia} = [\varpi + \varpi_{\rm ZP}]^{-1}$ and an error range 
$[D_{\rm min}$, $D_{\rm max}]$ based on the formal tabulated {\it Gaia} errors, $\varpi \pm \sigma_{\varpi}$.  
All comparisons between $D_{\rm Gaia}$ and $D_{\rm phot}$ assume that the ``true parallax distance" lies
between these bounds, with possible discrepancies arising from fluctuations in the parallax offset.   
Later in this paper, we explore possible dependences of the distance ratio, $D_{\rm Gaia}/D_{\rm phot}$,
on the relative parallax errors ($\sigma_{\varpi} / \varpi$), stellar distance, and SpT.    

We have avoided statistical corrections to parallax distances, because of the lack of a physical model to 
characterize the parallax offsets.  In this paper, we compare our photometric distances to offset-corrected
{\it Gaia}-DR2 distances and to two previous sets of photometric distances: 
an \IUE\ interstellar survey  of intermediate ions (Savage \etal\ 2001) and the \FUSE\ survey of \OVI\ in the 
Galactic disk (Bowen \etal\ 2008).   Differences in photometric distances between our new values and these 
surveys arise primarily from the updated spectral types and luminosity classes, which can change absolute 
magnitudes $M_V$ by 0.3--0.6 magnitudes.  Section~2 reviews previous absorption-line surveys with OB-star 
targets.  Section~3 describes the sample of OB stars and our techniques for deriving photometric and {\it Gaia}
parallax distances.   
Of special value in comparing $D_{\rm Gaia}$ to $D_{\rm phot}$ is a subset of 84 of the 139 stars with new 
classifications of spectral type and luminosity class from the GOS spectroscopic survey;  81 of these 84 stars
have reliable {\it Gaia} parallax distances.  For the other stars in our survey, we compute $D_{\rm phot}$ from 
SpTs and photometric data in the literature.   As illustrated in several figures, we find a wide dispersion in the 
ratio of photometric to parallax distances.   In Section~4 we apply our methods to two OB associations 
(Car OB1 and Per OB1) with potential changes to their historical distances.   On average, we find reasonable 
agreement between our photometric distances and $D_{\rm Gaia}$, but with considerable scatter in the ratio, 
particularly for stellar distances $d > 1.5$~kpc and parallax errors greater than 8\%.  We conclude with suggestions
for future applications, should {\it Gaia} analyses better characterize the parallax offsets.  This would allow us to 
calibrate the stellar classifications and provide more accurate distances to OB associations.

\section{INTERSTELLAR SURVEYS TOWARD GALACTIC OB STARS} 

With their high surface temperatures and far-UV continuum fluxes, massive OB-type stars provide bright
UV background sources for absorption-line surveys of Galactic interstellar gas (Spitzer \& Jenkins 1975;
Savage \& Sembach 1996).   Exploiting the strong UV resonance lines  of many elements, these studies have
quantified the gaseous content and spatial extent of  the ISM in atomic hydrogen (\HI), molecular hydrogen 
(H$_2$), and many heavy elements (e.g., C, N, O, Mg, Al, Si, P, S, Cl, Ar, Mn, Fe, Ni) over a range of ionization
states.   Of particular importance were the far-UV surveys of atomic and molecular hydrogen and of the \OVI\ 
doublet (1031.926~\AA\  and 1037.627~\AA) which identified a phase of hotter shock-heated interstellar gas at 
temperatures $10^5$~K to $10^6$~K.  Since the beginnings of ultraviolet space astronomy in the late 1960s, 
astronomers have employed a series of UV satellites with spectroscopic instruments to conduct gaseous 
abundance surveys using OB-stars as background continuum sources.  These satellites included 
{\it Copernicus}, \IUE,  \FUSE, and {\it Hubble Space Telescope} and measured absorption column densities, 
$N({\rm cm}^{-2}$), along several hundred stellar sight lines.  These column densities were translated to average 
number densities, $\bar{n} = N/d$, along the sight line to each target star, using estimates of its distance ($d$). 

The distances to these OB stars were usually photometric estimates from their apparent visual magnitudes ($V$),
corrected for extinction ($A_V$) and referenced to absolute magnitudes ($M_V$) inferred from the spectral type 
and luminosity class of the star.  As we discuss later, photometric distances come with considerable uncertainty, 
arising from errors in stellar photometry and extinction and from possible stellar mis-classification which affects 
absolute magnitudes.  Distances to the massive O-type stars are required to determine the luminosity density and 
ionizing photon fluxes in the Galactic disk and low halo (Dove \& Shull 1994;  Vacca \etal\ 1996).  Distances are 
also needed to compute the stellar luminosity, correlate the absorption with intervening gas and dust, and place 
the OB associations into the context of Galactic structure, spiral arms, and molecular clouds.  For example, 
from the Galactic latitude ($b$) of the stars and their distance above the disk plane, $z = d \sin b$, one can model
the vertical scale height of the gas layers (Savage \etal\ 1977; Shull \& Van Steenberg 1985; Diplas \& Savage 
1994b; Bowen \etal\ 2008).   

The Galactic ISM was first surveyed in the \Lya\ absorption line of atomic hydrogen by the OAO-2 satellite 
toward 69 stars of spectral type B2 and earlier, at average distances of 300 pc from the Sun (Savage \& Jenkins 
1972). The OAO-2 survey was later extended to 95 hot stars (Jenkins \& Savage 1974) and the {\it Copernicus}
 OAO-3 satellite measured \Lya\ absorption toward 100 OB stars within 1-2 kpc (Bohlin \etal\ 1978).  The
 {\it International Ultraviolet Explorer} (\IUE) was used for surveys of \HI\ toward 205 OB stars out to 5 kpc
 (Shull \& Van Steenberg 1985) and toward 554 hot stars with heliocentric distances up to 11 kpc (Diplas \& 
 Savage 1994a).  The latter survey was reduced to a working sample of 393 OB stars (Diplas \& Savage 1994b) 
 after excluding B1.5 and B2 stars contaminated by stellar \Lya\ absorption.  Interstellar molecular hydrogen
 (H$_2$) was surveyed in its lowest ($J =0$ and 1) rotational states of the far-UV Lyman and Werner bands 
 using data toward 109 OB-stars with {\it Copernicus} (Savage \etal\ 1977).   Two decades later,
 the {\it Far Ultraviolet Spectroscopic Explorer} (\FUSE) measured interstellar H$_2$ absorption toward hot
 OB-type stars and quasars (Shull \etal\ 2000; Browning \etal\ 2003).  \FUSE\ also surveyed H$_2$ along 38
 translucent lines of sight  to OB stars with visual extinction $A_V = 1.0-1.5$ (Rachford \etal\ 2002, 2009) and
 toward 70 OB stars in the Large and Small Magellanic Clouds (Tumlinson \etal\ 2002).  
Ultraviolet  satellites also conducted surveys of heavy elements, including studies of \CI\ with {\it Copernicus} 
(Jenkins \etal\ 1983)  and  the Space Telescope Imaging Spectrograph (STIS) on the {\it Hubble Space 
Telescope} (Jenkins \& Tripp 2001, 2011). The  \IUE\ surveyed low ionization states of \SiII, \MgII, \FeII, \SII,
and \ZnII\ (Van Steenberg \& Shull 1988) and intermediate ions \AlIII, \CIV, \SiIV (Savage \etal\ 2001).  The
highly ionized ISM phase was studied in \OVI, first with {\it Copernicus} (Jenkins \& Meloy 1974) and later 
with \FUSE\ (Bowen \etal\ 2008).   For our photometric distance calculations, we employ two grids of absolute 
magnitudes:  our standard OB-star grid from Bowen \etal\ (2008) and an alternative O-star grid from 
Martins \etal\ (2005).   

\section{DISTANCES FROM PARALLAXES AND PHOTOMETRY}

\subsection{The Stellar Sample}

The 139 OB-type stars chosen for our current study served as UV-background sources for \FUSE\ 
observations of \Htwo\ absorption in the diffuse ISM.  A survey of \Htwo\ column densities in various 
rotational states of the ground vibrational state (J. M. Shull \etal\ 2019, in preparation) uses these OB 
stars as bright targets ($V <  10$) with typical color excesses $E(B-V) < 0.5$.    To update photometric 
distances to these stars, we have conducted an extensive review of the literature for SpTs and photometry:
$B$ and $V$ magnitudes.  The color excess $E(B-V)$ was derived from $(B-V)$ relative to intrinsic 
colors $(B-V)_0$, and visual extinction followed from $A_V = R_V E(B-V)$, using a standard value 
($R_V = 3.1$) for the ratio of total-to-selective extinction.   In Section 4.1, where we analyze distances 
to 29 O-type stars in the Carina Nebula, we adopt a higher value, $R_V = 4.0$, observed by  
Feinstein \etal\ (1973) and Tapia \etal\ (2003) and confirmed here by comparing $A_V$ to $E(B-V)$ for 
these stars.   

Table 1 gives the star names, along with internal ID numbers, Galactic coordinates, spectral types 
(SpT), and values of $B$, $V$, $E(B-V)$, and $A_V$.   Papers that we used for SpT and photometry
include classic studies by Morgan \etal\ (1955), Hiltner (1956), Hiltner \& Johnson (1956), Lesh (1968); 
Schild \etal\ (1969); Hill (1970),  Hill \etal\ (1974), Garrison \etal\ (1977),  Wesselius \etal\ 1982, and
Schild \etal\ (1983).  A full list of references is provided in footnote (b) of Table 2.  A valuable 
sub-sample of our 139 stars comes from 84 O-type stars with new spectral classifications from the 
GOS spectroscopic survey.  The main goals of GOS were to obtain high-S/N, moderate-resolution
($R \sim 2500$) blue-ultraviolet spectra of over 1000 O-type stars in the Milky Way and to derive 
spectral types classified according to well-defined standards (Walborn \etal\ 2010; Ma\'iz Apell\'aniz 
\etal\ 2011).  The GOS survey provides both photometry and stellar classification. In our Table~1, we 
use values ($V_{J,0}$ and $A_{V_J}$) from Ma\'iz Apell\'aniz \& Barb\'a (2018) who modeled optical 
and NIR photometry with their new family of extinction laws (Ma\'iz Apell\'aniz \etal\ 2014).

\subsection{Gaia Parallax Distances}

We began our survey by obtaining the basic stellar data of parallaxes and quoted errors (in milli-arcsec) 
from the on-line {\it Gaia}-DR2 archive.  The catalogue was queried on the {\it Gaia} archive at the website
\url{http://gea.esac.esa.int/archive}.   As recommended by the {\it Gaia} Mission Team (Lindegren \etal\ 2018;
Arenou \etal\ 2018) we have applied a constant parallax offset of 0.03 mas to these values, relative to
the International Celestial Reference Frame (ICRF) provided by a half-million quasars with accurate 
VLBI positions (Mignard \etal\ 2018).  The recent astronomical literature contains analyses of parallax 
offsets ranging from 0.029--0.081 mas for different stellar types (Cepheids, eclipsing binaries, 
red giants).  Riess \etal\ (2018) determined a 0.046 mas mean offset ($46 \pm 13 \;\mu$as) from their 
\HST\ sample of 50 Milky Way Cepheids, and they noted the apparent dependence of the offset on stellar 
magnitude, color, and position on the sky (Lindegren \etal\ 2018).   Stassun \& Torres (2018) found a 
mean offset of 0.082~mas ($82 \pm 33~\mu$as) for 89 eclipsing binaries.   However, a more recent study 
(Graczyk \etal\ 2019) of 81 detached eclipsing binaries found a lower mean offset, $0.031 \pm 0.011$~mas, 
in agreement with the value recommended by the {\it Gaia} team.  Zinn \etal\ (2019) found a mean offset 
of 0.053~mas  ($52.8 \pm 2.4 \; {\rm [rand]} \pm 8.6 \; {\rm [syst]}~\mu$as) using red-giant stars in the
{\it Kepler} field with well-characterized asteroseismic data.  

Several recent studies have employed statistical methods (Bailer-Jones \etal\ 2018; Davies \& Beasor 2019)
to translate {\it Gaia} parallax angles and their error range to the most-probable distance.  These studies 
emphasize the need to control for distance bias using statistical methods.  Because we are studying individual 
OB stars, in which the source of the offsets is poorly known, we believe such procedures are not well-suited
to our survey.  We follow a different procedure, using the {\it Gaia}-DR2 parallax and error, 
$\varpi \pm \sigma_{\varpi}$, to find a formal ``parallax distance", $D_{\rm Gaia} = [\varpi + \varpi_{\rm ZP}]^{-1}$, 
and the corresponding error range, after applying a standard zero-point offset $\varpi_{\rm ZP} = 0.03$~mas.   
We then compare $D_{\rm Gaia}$ to the
photometric distances and examine the differences and their dependence on stellar parameters, stellar 
distances, and relative parallax errors.    Our comparison of parallax distances with photometric distances
finds considerable differences when parallax errors are greater than 8\%.  As noted earlier, we used a standard
parallax offset of 0.03~mas, which we believe to be more appropriate for the blue colors of OB stars rather than 
higher offsets (0.05 mas) for red giants.  Table 2 lists the offset-corrected {\it Gaia} parallax distances for 135 of 
our 139 survey stars, together with the inferred range of distances based on quoted DR2 parallax errors.   We 
could not use {\it Gaia} data for four stars (\#79, \#80, \#98, \#119) owing to negative parallaxes or unacceptably
large errors.   We do not include possible systematic errors based on the dependence of offsets on brightness, 
color, or position on the sky.  Typical DR2 parallax errors for the best  data are $\pm0.03$~mas, comparable to 
the 0.03~mas applied offset.   For context, a star at 2.5~kpc distance has a parallax of 0.40 mas.   Therefore, 
offsets of 0.03--0.05~mas can produce 8-15\% fractional errors at typical 1-3~kpc distances to the OB stars 
in our survey.  

\subsection{Photometric Distances}

The initial rationale for this paper was a revised set of photometric distances toward the 139 OB stars used 
as background targets in our \FUSE\ survey of interstellar \Htwo.   Many of these stars appeared in previous 
surveys of interstellar matter.  For example, 100 stars are in common with the \IUE\ survey of \AlIII, \SiIV, \CIV\
by Savage \etal\ (2001), and 101 stars are in common with the \FUSE\ survey of \OVI\ in  the Galactic disk 
(Bowen \etal\ 2008).  Many of our 139 stars were used in past UV surveys of interstellar \HI\ (Shull \& Van 
Steenberg 1985; Diplas \& Savage 1994a,b) and heavy elements (Van Steenberg \& Shull 1988; Jenkins 2009).    
The new photometric distances in this paper were computed from the usual expression,
\begin{equation}
   D_{\rm Shull} = (10~{\rm pc}) \cdot 10^{ (V - A_V - M_V) /5 }   \; \;  ,
\end{equation} 
with absolute magnitudes $M_V$ from the standard grid in Bowen \etal\ (2008).  For all 139 stars in our study, 
we used critically evaluated photometry and extinction from the literature.


\begin{figure}
\includegraphics[angle=0,scale=0.6] {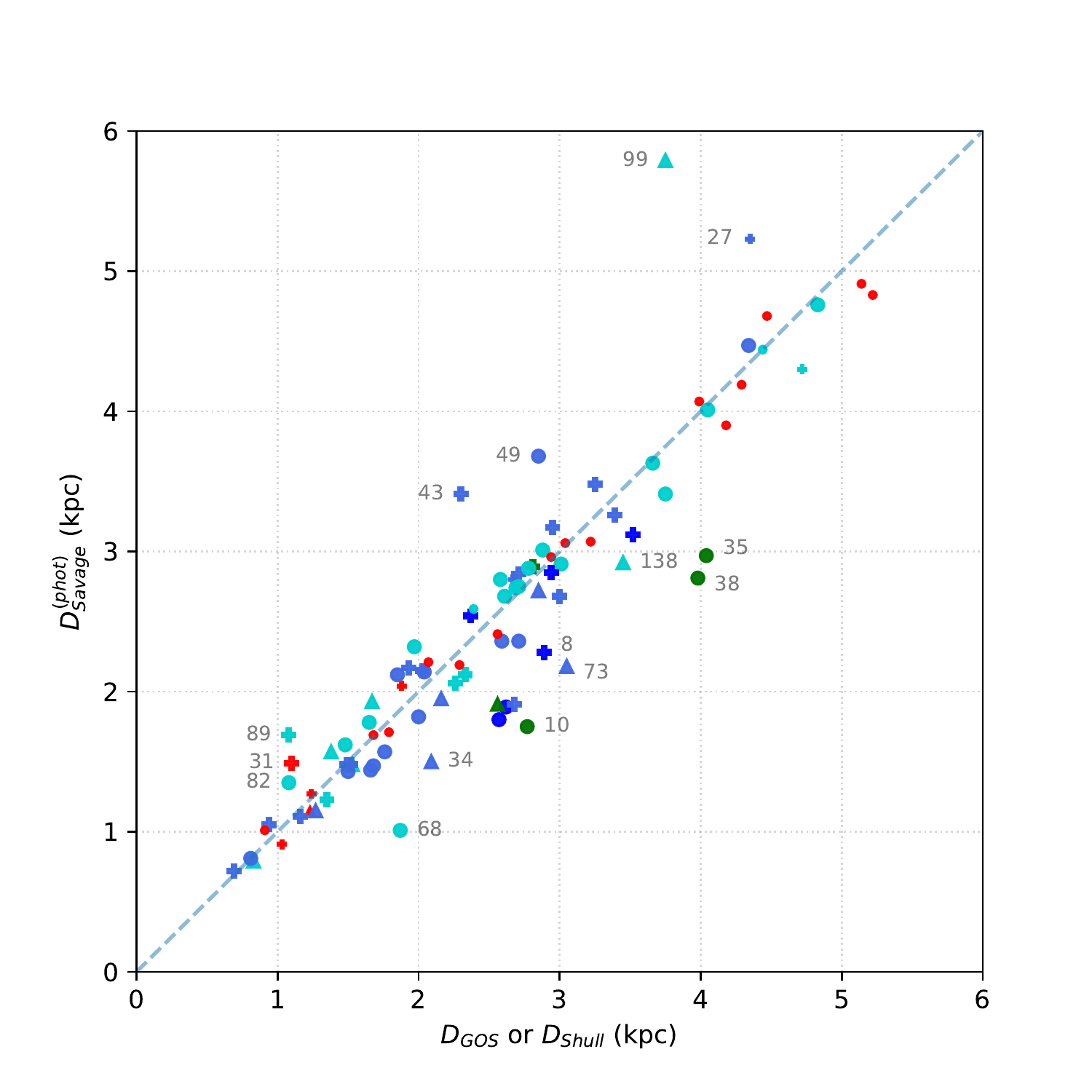} 
\caption{\small{ Comparison of  our new photometric distances, $D_{\rm Shull}$ or $D_{\rm GOS}$ when 
available, with $D_{\rm Savage}$ (Savage \etal\ 2001) for 100 OB stars in common with the \IUE\  survey.
Outliers are labeled by ID numbers (Tables 1 and 2) and discussed in Appendix~A.  Neither survey includes 
error bars, which are primarily systematic (assumptions on SpT and $M_V$).  Stars of similar spectral types 
are color-coded as follows:  
dark blue (O2--O4); cornflower blue (O5--O7); cyan = (O8--O9); dark green (ON and WN); red (B0--B4). 
Luminosity classes are shown as follows:  circles (supergiants I, Ia, Iab, Ib, II); triangles (giants II-III, III, III-IV, IV); 
crosses (main sequence V, IV-V, and unknown).  }
 }
\end{figure}


Potentially more accurate photometric distances may be found for 84 of our 139 stars that appear in the 
GOS spectroscopic survey (Ma\'iz Apell\'aniz \etal\ 2004).  Revised spectral types for these GOS stars 
were provided by Sota \etal\ (2011, 2014), and digital stellar photometry and extinction were tabulated by 
Ma\'iz Apell\'aniz \& Barb\'a (2018).   Based on optical and near-infrared photometry, they list the 
extinction-corrected visual magnitude, $V_{J,0} \equiv V_J - A_{V_J}$, where the visual extinction $A_{V_J}$ 
was derived from a new family of extinction laws (Ma\'iz Apell\'aniz \etal\ 2014).   
We denote the photometric distances for the GOS survey stars by 
\begin{equation}
   D_{\rm GOS} = (10~{\rm pc}) \cdot 10^{ (V_{J,0} - M_V) / 5}   \; \;  .
\end{equation} 
It is important to determine appropriate values of $M_V$,  the star's absolute magnitude derived from the 
star's spectral type (SpT) and luminosity class.  Accurate photometric distances require calibration 
of stellar type (for $M_V$) as well as visual magnitude $V_{J}$ and extinction $A_{V_J}$.


\begin{figure}
\includegraphics[angle=0,scale=0.6] {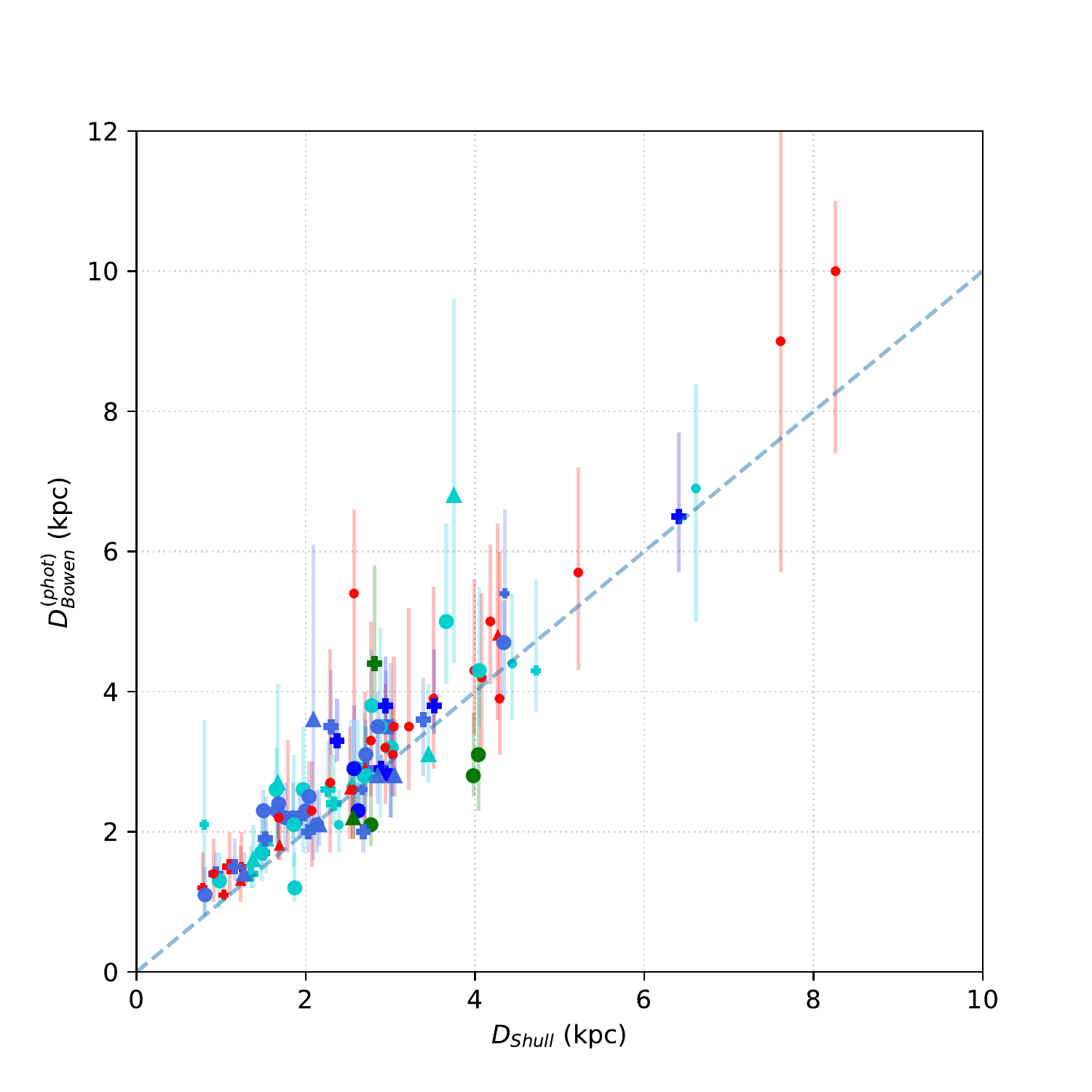} 
\caption{\small{ Comparison of new photometric distances, $D_{\rm Shull}$ in Table 2, with 101 OB stars
in common with the \FUSE\ survey of \OVI\ in Galactic disk (Bowen \etal\ 2008).  Vertical error bars were 
supplied by Bowen \etal\ (2008).  Color-coding and symbols are the same as in Fig.\ 1.  } 
 }
\end{figure}


Table 2 lists photometric distances tabulated in two previous ISM surveys, denote as $D_{\rm Savage}$ 
from the \IUE\ survey (Savage \etal\ 2001) and $D_{\rm Bowen}$ from the \FUSE\ survey (Bowen \etal\ 2008).  
We also list the offset-corrected parallax distance ($D_{\rm Gaia}$) and two photometric distance calculations 
from our current study, denoted $D_{\rm Shull}$ and $D_{\rm GOS}$ (see eqs.\ [1] and [2]).  Figures 1 and 2 
compare our new photometric distances with values from previous surveys, $D_{\rm Savage}$ and 
$D_{\rm Bowen}$.   We find reasonable agreement with $D_{\rm Savage}$ in most cases ($d < 5$~kpc) 
as seen by the scatter about the one-to-one ratio line.   We label 15 outliers on Figure 1, for which the
distances differ by more than 15-20\%.   Deviations from Bowen \etal\ (2008) are larger (Figure 2) because
of differences in photometry ($V$) and extinction $E(B-V)$.   In general, the changes in distance arise from 
the updated spectral types, particularly luminosity classes which can change the absolute magnitudes by 
$\Delta M_V = 0.3-0.6$ magnitudes (factors of 1.15--1.32 in distance).  In several cases, changes in the 
visual extinction contribute to the discrepancies.   Appendix~A discusses the outliers in Figure~1 and other 
stars with SpT changes.  The subset of 84 GOS stars is particularly useful because their updated stellar 
classifications and digital photometry provide the extinction-corrected visual magnitude, 
$V_{J_0} \equiv V_J - A_{V_J}$.  The photometric distances labeled $D_{\rm GOS}$ require no assumptions 
about $R_V$, provided that one trusts the GOS extinction model (Ma\'iz Apell\'aniz \etal\ 2014).


\begin{figure}
\includegraphics[angle=0,scale=0.6] {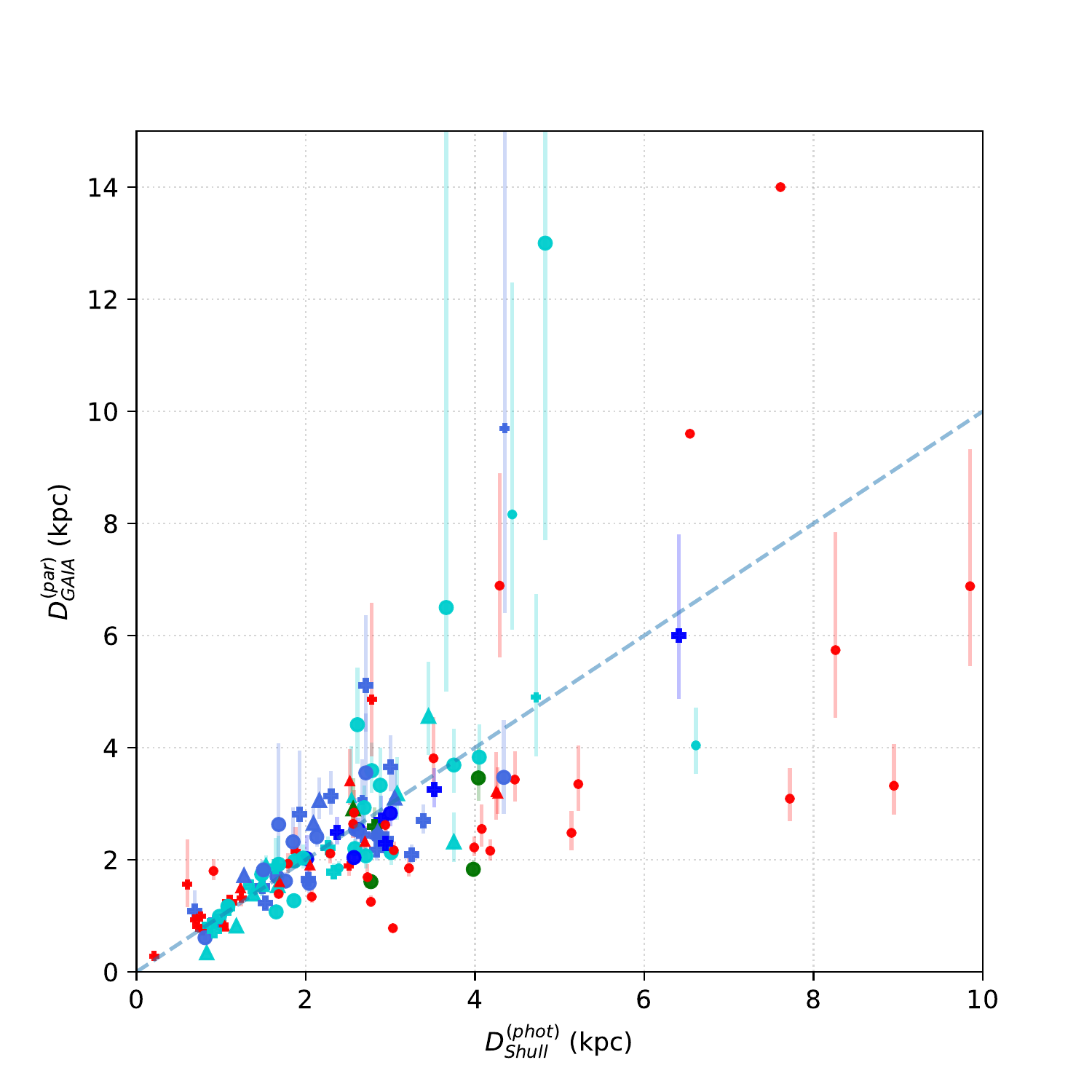} 
\caption{\small{ Comparison of new photometric distances, $D_{\rm Shull}$, with offset-corrected parallax 
distances ($D_{\rm Gaia}$) for 135 of the 139 OB stars in our survey.   Four stars had missing or unusable
{\it Gaia} data.  Color-coding and symbols are as in Fig.\ 1.  Increasingly large discrepancies appear at 
distances $D  >  1.5$~kpc, particularly for the  B-type stars (red symbols).  } 
 }
\end{figure}


Figure 3 compares our photometric distances with {\it Gaia} distances for 135 stars of the 139 stars. 
The vertical error bars on $D_{\rm Gaia}$ reflect the internal uncertainties listed in the DR2 database.
We have not plotted errors on photometric distances, which are primarily systematic uncertainties
in SpT and $M_V$.  Evidently, the {\it Gaia}-DR2 parallax distances track photometric distances out
to distances $d \approx 1.5$~kpc.   Increasing scatter and large discrepancies appear at $d  >  1.5$~kpc, 
particularly for early B-stars (red symbols).  Within the accuracy of the data, it is difficult to evaluate whether 
the standard (0.03~mas) parallax offset is any better than higher values (0.05~mas) found for other sources.   
With the next {\it Gaia} data release, it may be possible to evaluate the appropriate offset.


\begin{figure*}
\includegraphics[angle=0,scale=1.18] {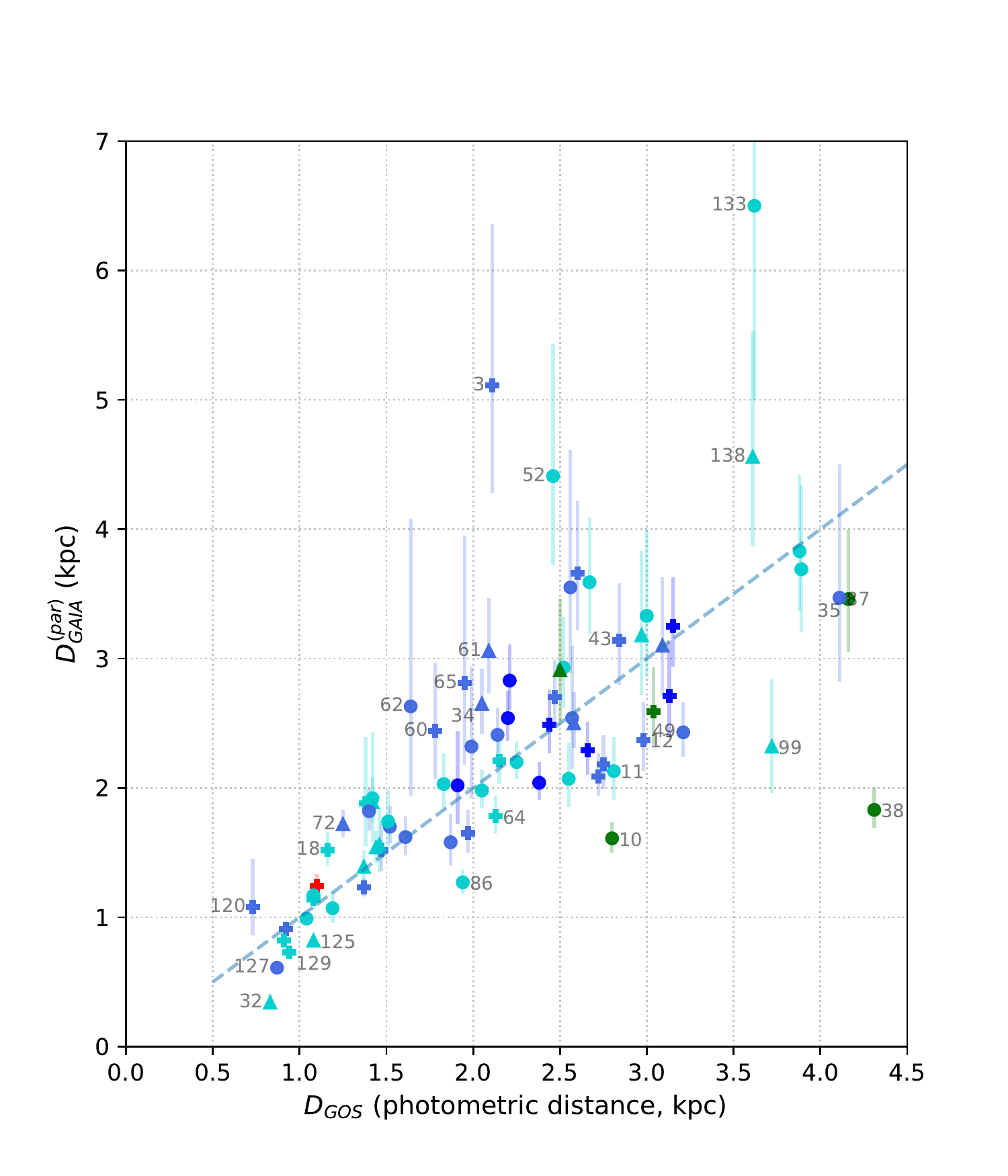} 
\caption{\small{Comparison of {\it Gaia}-DR2 parallax distances and photometric distances for a 
 subset of 84 O-type stars with new (GOS) stellar classifications (Sota \etal\ 2011, 2014) and digital 
 photometry (Ma\'iz Apell\'aniz \& Barb\'a 2018).  Three of the 84 stars have unusable {\it Gaia} parallaxes.  
 Outliers are labeled with their ID numbers (Tables 1 and 2).  Note the increasing dispersion about the
 dotted line of equality at distances $d > 1.5$~kpc.  More outliers lie above the line 
 ($D_{\rm Gala} > D_{\rm GOS}$) suggesting a broad, asymmetric distribution in parallax offsets.   
Color-coding and symbols are as in Figure~1.  Star \#31 with $D_{\rm GOS} = 1.10$~kpc is labeled with 
a red cross (for B0.2~V) although it was classified as O9.7~II by GOS; see Appendix~A.     } 
}

\end{figure*}


Figure~4 provides a similar comparison of $D_{\rm Gaia}$ with photometric distance, $D_{\rm GOS}$,
for the subset of 81 GOS stars with useful parallaxes.  In principle, these stars should have more reliable 
distances, owing to their updated SpTs, digital photometry, and modeled extinction ($A_V$).  However, 
we continue to see differences between $D_{\rm Gaia}$ and $D_{\rm GOS}$ at $d > 1.5$~kpc.
We label the outliers with their ID numbers from Tables 1 and 2 and provide a detailed discussion of
the stellar properties and possible reasons for the discrepancies in Appendix A.  Four of these stars 
(\#10, \#35, \#38, \#99) also appeared as outliers on Figure~1.  In the case of star \#99 (HD~168941), 
our new photometric distance, $D_{\rm GOS} = 3.72$~kpc, and the parallax distance,
$D_{\rm Gaia} = 2.32$~kpc, are considerably less than previous photometric distances of 6.1~kpc 
(Tripp \etal\ 1993) and 5.79~kpc (Savage \etal\ 2001).  The difference hinges on the star's correct  
luminosity class (IV vs.\  II-III).   
 
Figure 5 provides further insight into the Figure 4 outliers.  Using data (Table 3) on the parallax angle
($\varpi$) and its formal error ($\sigma_{\varpi}$) from {\it Gaia}-DR2 we plot the parallax-to-photometric 
distance ratio, $D_{\rm Gaia} / D_{\rm GOS}$, versus $(\varpi / \sigma_{\varpi}$),  an indicator of parallax 
quality.  Although most of the GOS stars lie within $\pm30$\% of the line of equality, 
20 outliers with $D_{\rm Gaia} > 1.3 D_{\rm GOS}$ (top-left corner) have parallax errors exceeding 20\%. 
Six outliers with $D_{\rm Gaia} < 0.7 D_{\rm GOS}$ (bottom-right corner) have parallax errors less 
than 10\%.  These include stars \#38, \#10, \#86, and \#127, each discussed in Appendix A.  
It is surprising that these stars with low parallax errors would differ from the photometric distances.  
Some of these discrepancies could arise from complications of close binary orbits on parallax
measurements.  The 20 stars with $D_{\rm Gaia} > 1.3 D_{\rm GOS}$ have mean {\it Gaia} magnitudes 
$\langle G \rangle = 7.42$, similar to the mean of all 81 GOS stars, $\langle G \rangle = 7.55$.   
Figure~5 suggests that reliable parallax distances require {\it Gaia} errors less than about 8\%.


\begin{figure*}
\includegraphics[angle=0,scale=1.05] {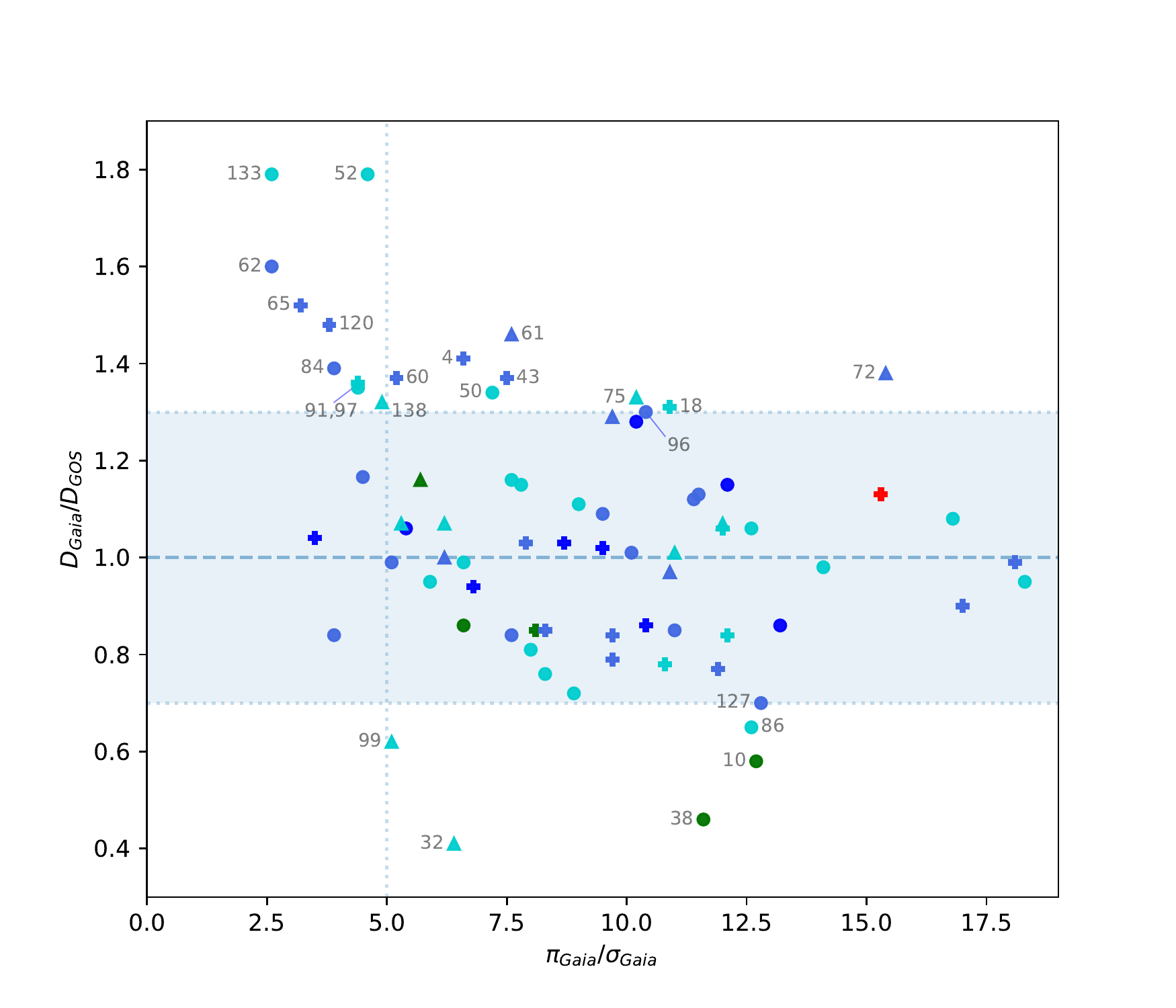} 
\caption{\small{We plot the distance ratio ($D_{\rm Gaia} / D_{\rm GOS}$) vs.\ $(\varpi / \sigma_{\varpi})$, a quality 
indicator of {\it Gaia} parallax measurements (see Table 3).   Color-coding and symbols are as in Figure~1. 
Shaded band encloses the boundaries of $\pm30$\% deviations from equality, and vertical dotted line at 
$(\varpi / \sigma_{\varpi}) = 5.0$ marks 20\% parallax errors.   Of  81 GOS O-type stars with reliable {\it Gaia} 
parallaxes, 26 lie outside the $\pm30$\% band in an asymmetric distribution:  20 stars have
$D_{\rm Gaia} / D_{\rm GOS} \geq 1.30$ and 6 stars have $D_{\rm Gaia} / D_{\rm GOS} \leq 0.70$.
Two stars lie above the top of the plot:  
Star \#3 with $D_{\rm Gaia} / D_{\rm GOS} = 2.42$ ($\sigma_{\varpi} / \varpi = 0.23$) and 
Star \#70 with $D_{\rm Gaia} / D_{\rm GOS} = 2.66$ $(\sigma_{\varpi} / \varpi = 1.28$).  Reasonable 
distance agreement ($D_{\rm Gaia} \approx D_{\rm GOS}$) seems to require parallax errors less than 8\%.
The outliers, labeled with ID numbers and discussed in Appendix A, may reflect a wide, asymmetric 
distribution of parallax offsets about the mean value of 0.03~mas.  Some of the discrepancies may be
produced by binary effects or incorrect luminosity classes (GOS spectroscopic survey). } 
}

\end{figure*}

\section{APPLICATION TO TWO OB ASSOCIATIONS}  

In addition to comparing individual photometric and parallax distances for 139 OB-type stars in our survey,
we applied our techniques to associations of stars.   From our sample of O-type stars, supplemented by other 
O-stars in the GOS survey, we used mean values of {\it Gaia} parallax distances and new photometric distances 
to estimate distances to two well-known OB associations: Perseus OB1 and Carina OB1.   For this comparison, 
we used two photometric distance estimates, $D_{\rm Shull}$ and $D_{\rm GOS}$, for selected O-stars.  
Table 4 shows the results for 29 O-stars in Carina OB1, and Table 5 lists data for 12 O-stars in Perseus~OB1.   
Both OB associations have been studied extensively, but their estimated distances span a wide range with 
historical disagreements over their connection with nearby (or embedded) star clusters: 
the  famous ``Double Cluster" $h$ and $\chi$ Persei (Johnson \& Morgan 1955;  Slesnick \etal\ 2002) near 
Per~OB1; and the clusters Trumpler 14, Trumpler 16, Trumpler 15, and Collinder 228 in the Carina Nebula 
(Humphreys 1978;  Massey \& Johnson 1993). 

\subsection{Carina OB1 star clusters}  

Historical controversy exists (Davidson \& Humphreys 1997; Walborn 2012) over the distances to Car~OB1 
and its associated star clusters, Trumpler~14, Trumpler~15,Trumpler~16, Collinder~228.  Th\'e \& Vleeming 
(1971) derived distances of 2.0~kpc (Tr~14) and 2.5 kpc (Tr~16) and suggested a distance of $2.5\pm0.2$~kpc 
to the $\eta$~Carinae Nebula.  From a small number of O-type stars,  Walborn (1973b) found DM = 12.72  (Tr~14),  
12.11 (Tr~16), and 12.18 (Col~228).   He concluded that Tr~16 and Col~228  ``form a single, very young complex 
located at a distance of 2600~pc" and that  Tr~14 ``is an exceedingly young, compact cluster which may be as 
distant as 3500~pc".   After considering age differences of the clusters,  Walborn (1982) later revised these 
distances to a common value, DM = $12.26 \pm 0.12$ ($2.83 \pm 0.16$~kpc).  
Feinstein \etal\ (1973) estimated DM =  $12.65 \pm 0.20$ ($3390 \pm 300$~pc) assuming that Tr~14 and 
Tr~16 form a common group. Initially adopting $R_V = 3.0$, they corrected the distance to 2650~pc with
 $R_V \approx 4.0$ because the ($V-I$, $B-V$) array indicated anomalous extinction.  They also noted a 
 close relationship between the emission nebula, dust, and stars in the cluster.  Humphreys (1978) adopted a 
distance modulus DM = 12.7 (3.5 kpc) for Tr~14 and DM = 12.1 (2.6 kpc) for Tr~16, assuming $R_V = 3.0$.   
Using near-infrared (JHKL) photometry, Tapia \etal\ (1988) adopted $D = 2.4 \pm 0.2$~kpc, for 
Tr~14, Tr~15, Tr~16, and Coll~228.   Massey \& Johnson (1993) found DM = $12.55 \pm 0.08$ (3.2~kpc) for 
early-type stars in Tr~14 and Tr~16 with $R_V = 3.2$. 
 
Many of the differences in these photometric distances arise from the adopted extinction law and the choice 
of $R_V = A_V/E(B-V) = 3.0$, 3.2, or 4.0.   Evidence of anomalous extinction in Carina ($R_V = 4.0$) was first
suggested by Feinstein \etal\ (1973).  From their $UBVI$ CCD photometry, Hur \etal\ (2012) also found 
abnormal reddening with $R_V = 4.4 \pm 0.2$ for stars in the $\eta$~Carinae Nebula and concluded that
Tr~14 and Tr~16 have practically the same DM = $12.3 \pm 0.2$ ($2.9 \pm 0.3$~kpc).  Tapia \etal\ (2003) 
carried out large-scale imaging ($UBVRIJHK$) of the Carina Nebula and found DM  $= 12.14 \pm 0.67$, 
with large scatter in both $A_V$ and distance.   For the individual clusters, they found mean distance moduli 
of $12.23 \pm 0.67$ (Tr~14) and $12.02 \pm 0.57$ (Tr~16).  The Tr~14 and Tr~16 clusters are now regarded
to have similar distances, and associated with the massive, eruptive star $\eta$ Carinae in Tr~16.   
Spectroscopic velocities of the ejected filaments of the Homunculus Nebula in $\eta$~Car, combined with 
the estimated time of ejection, indicated distances of $\sim2.5$~kpc  (Hillier \& Allen 1992) and $2250 \pm 180$~pc
(Davidson \etal\ 2001).  Davidson \& Humphreys (1997) previously estimated the distance to $\eta$~Car 
at $2.3 \pm 0.2$~kpc based on luminosities of O stars in Tr~16 and expansion of the $\eta$~Car ejecta.   
Using similar methods, Smith (2006) found $D = 2350 \pm 50$ kpc from proper motions of the 
Homunculus Nebula.   The small error on this distance may not include systematic errors in the geometric 
assumptions. 

In our study, we used data toward 29 O-type stars in the Carina Nebula, located in four star clusters.  
Our statistical analysis (Table 4) considered 16 stars in Tr~16,  nine stars in Tr~14, three stars in Coll~228, 
and one star in Tr~15.   We did not include HD~93206 (QZ~Car), a complex double-binary system 
(ID \#46 in Appendix A).   We compared three estimates of photometric distance (columns 8, 9, 10 in Table~4) 
with {\it Gaia} parallax distance (column 7).    The photometric distances were evaluated in three ways.  
Distances $D_{\rm phot}$ are based on photometry corrected for extinction, similar to $D_{\rm Shull}$ in 
Table~2 but with an anomalous extinction law, $R_V = 4.0$, rather than the standard value of $R_V = 3.1$.  
Distances labeled $D_{\rm GOS}$ adopt values of $V_{J,0}$ and $A_{V_J}$ from Ma\'iz Apell\'aniz \& 
Barb\'a (2018) and listed in columns 3 and 4 of Table~4.   These distances are based on our standard grid 
of absolute magnitudes (Bowen \etal\ 2008).  Distances labeled $D'_{\rm GOS}$ follow the same procedure 
with GOS photometry, but employ the $M_V$ grid of Martins \etal\ (2005).   For the four clusters in Car~OB1, 
we find mean distances $\langle D_{\rm GOS} \rangle$ of $2.55\pm0.30$~kpc (Tr~16),  
$2.68\pm0.31$~kpc (Tr~14),  and $2.58\pm0.35$~kpc (Coll~228).  The single O-star in Tr~15 had a 
distance of 2.76~kpc.   

We find no significant difference in distances among the four clusters.  For the ensemble of all 
29 O-type stars, we find mean photometric distances 
$\langle D_{\rm GOS} \rangle = 2.60 \pm 0.28$~kpc (absolute magnitudes of Bowen \etal\ 2008) and
$\langle D'_{\rm GOS} \rangle = 2.42 \pm 0.29$~kpc (absolute magnitudes of Martins \etal\ 2005).  
The mean parallax distance is $\langle D_{\rm Gaia} \rangle \approx 2.87 \pm 0.73$~kpc.    
Davidson \etal\ (2018) suggested that the O3 stars in Carina OB1 provide a special population,
because of their young age and likely formation proximity.  For Tr~16, they noted four such stars 
(HD~93205, HD~93250, HD~303308, and MJ~257) whose (uncorrected) {\it Gaia}-DR2 parallaxes 
had a small dispersion, $\langle \varpi \rangle = 0.383 \pm 0.017$~mas.  After applying a 0.030~mas 
offset, this corresponds to parallax distance $\langle D_{\rm Gaia} \rangle = 2.42^{+0.19}_{-0.16}$~kpc.   
Our survey includes three of these stars (ID \#45, \#48, and \#137 in Table 4) whose updated GOS 
spectral types are O3.5~V, O4~III, and O4.5~V, respectively.  

\subsection{Perseus OB1}  

The two open clusters $h$ and $\chi$ Persei (the Double Cluster in Perseus) have appeared in the
literature (Garmany \& Stencel 1992) with distance moduli ranging from DM = 11.4--12.0 corresponding
to 1.9--2.5 kpc.   Schild (1967) placed $h$~Persei 350~pc more distant and 5 Myr older than $\chi$~Persei, 
with estimates of 2.15~kpc ($h$) and 2.50~kpc ($\chi$) and an association of ``outer group" stars at 
intermediate distances.   In a CCD~UBV imaging survey, Slesnick \etal\ (2002) found nearly identical distance 
moduli, $11.85 \pm 0.05$ ($2.34 \pm 0.05$ kpc) for stars near the cluster nuclei.  However, they did not 
resolve the question of whether the double cluster is located at the core of Per~OB1.   
Currie \etal\ (2010) used photometric and spectroscopic observations of stars in $h$ and $\chi$ Persei,
finding nearly identical properties and distance moduli 11.80--11.85 (2.29--2.34~kpc).  Zhong \etal\ (2019) 
used {\it Gaia}-DR2 data to suggest filamentary substructure extending 200~pc away from the Double Cluster.
Also using {\it Gaia}-DR2 data on red supergiants, Davies \& Beasor (2019) found a distance of 
$2.25^{+0.16}_{-0.14}$ kpc for $\chi$~Persei.   Lee \& Lim (2008) noted the bulk motions of luminous 
members of this association away from the Galactic plane and the absence of any giant molecular cloud 
in its vicinity. They suggested sequential star formation in a shell of molecular gas pushed outward by an 
expanding superbubble.   In fact, the O-stars in Per OB1 are spread over $\sim6-8^{\circ}$ (Humphreys 
1978) corresponding to 250-320 pc and consistent with the dispersion in distances.   The early O-type stars 
are likely to be younger than the supergiants. The latter are spread across several degrees of longitude 
with a corresponding dispersion in distances. 

Using data for 12 O-type stars in Per~OB1 with GOS photometry, including six from Table~2 and six others
(see Table~5),  we compared the new photometric distances ($D_{\rm Shull}$ and $D_{\rm GOS}$) and the 
{\it Gaia} parallax distance ($D_{\rm Gaia}$).   From this list, we excluded three stars:  HD~15642 and 
HD~14442 (discrepant large distances) and HD~14633 (at lower Galactic latitude) which are considered
uncertain or unlikely association members (Lee \& Lim 2008).  For the remaining 9 stars, we found mean 
distances of
$\langle D_{\rm phot} \rangle = 2.95 \pm 0.23$~kpc, $\langle D_{\rm GOS} \rangle = 2.99 \pm 0.14$~kpc, 
and  $\langle D_{\rm Gaia} \rangle = 2.47 \pm 0.57$~kpc.   One possible cause of the difference could 
be that our standard grid of absolute magnitudes (Bowen \etal\ 2008) is too luminous.  Adopting the 
Martins \etal\  (2005) grid of $M_V$, we found a 7\% lower mean distance, 
$\langle D'_{\rm GOS} \rangle = 2.77 \pm 0.22$~kpc.   However, this distance was based on only seven
stars whose SpTs could be matched or interpolated on the Martins \etal\ (2005) grid.  Unfortunately, they
only list three luminosity classes (V, III, I) with large jumps in $M_V$ for classes II and IV.  

\section{SUMMARY AND FUTURE DIRECTIONS}

With the availability of new O-star spectroscopic surveys, {\it Gaia}-DR2 parallaxes, and technical advances
in modeling stellar atmospheres and evolutionary tracks, it is both appropriate and timely to re-assess 
basic parameters for the most massive stars in our Galaxy.   These include photometric distances, which
depend on the absolute magnitudes ($M_V$) associated with SpT and luminosity class.  We have calculated
new photometric distances to 139 OB-type stars and compared them to parallax distances from {\it Gaia}-DR2,
applying a standard (0.03 mas) parallax offset from the quasar celestial reference frame.  
Of special value were 84 stars from the GOS spectroscopic survey (Ma\'iz Apell\'aniz \etal\ 2004) which 
generated a large sample of O stars within several kpc of the Sun with updated spectral types
(Sota \etal\ 2011, 2014),  accurate digital photometry, and corrections for optical-NIR dust extinction 
(Ma\'iz Apell\'aniz  \& Barb\'a 2018).   We used GOS information for these stars, but also compiled values 
of critically evaluated photometry from the literature for all 139 stars.   The GOS stars are presumed 
to provide more reliable photometric distances, owing to updated spectral types, digital photometry, and 
and reddening corrections.   However, as discussed in Appendix~A, we explored possible reasons for the
discrepancies between photometric and parallax distances for the outliers on Figures 1, 4, and 5.   In most 
cases, the differences result from changes in SpT and luminosity class.  These outliers present an opportunity 
to assess whether large parallax offsets or incorrect stellar classification explain the differences.

A sizable fraction ($\sim30$\%) of the stars in Figures 4 and 5 exhibit significant differences between our 
photometric distances and those derived from {\it Gaia}, particularly at $d > 1.5$~kpc and when {\it Gaia} 
parallax errors exceed 8\%.  The ratio of photometric-to-parallax distances exhibits increasingly large 
fluctuations about the unit-ratio line in Figure 4.   Figure~5 shows that some but not all of these discrepant 
stars have large parallax errors ($\sigma_{\varpi} / \varpi > 0.08$).     
In addition to possible SpT uncertainties, there are also likely systematic errors in {\it Gaia}-DR2 parallaxes.  
Stars above the slope-one line would typically require parallax offsets of  $+0.1$~mas  to 0.2~mas to bring
outliers into agreement ($D_{\rm Gaia} \approx D_{\rm GOS}$).  Stars below the dotted line would require
comparable {\it negative} parallax offsets, although there are fewer of such stars.  Four of these stars 
(\#10, \#38, \#86, \#99) would require offsets of  $-0.08$~mas to $-0.23$~mas, instead of the standard 
$+0.03$~mas.    Alternatively, their updated SpTs from GOS may be incorrect, particularly the luminosity
classes.  Reconciling these distances will require careful examination of the GOS classifications, 
together with improved parallax measurements.   

We have not listed errors in photometric distances, which arise primarily from systematic uncertainties in 
photometry ($B$ or $V$ magnitudes), extinction corrections ($A_V$), and the adopted grid of absolute 
magnitudes ($M_V$).  For the 84 GOS stars (Figures 4 and 5) the photometry ($V_J$ and $A_V$) are 
typically accurate to $\pm0.02$ magnitudes.  Thus, changes in SpT or luminosity class are the primary
source of differences in our revised photometric distances compared to previous values 
(Savage \etal\ 2001;  Bowen \etal\ 2008).   A few discrepancies (Figures 1 and 2) arise from different choices
of $E(B-V)$ and from using $A_V$ from the GOS survey.  In some of those cases, such as stars in the Carina 
Nebula, we found evidence for  anomalous dust extinction ($R_V \approx 4$).

\vspace{0.4cm}

\noindent
We now summarize the major results of our  survey:
\begin{enumerate}

\item We find reasonable agreement in our new photometric distances with prior values (Savage \etal\ 2001) 
for many of the 100 OB stars in common.  Several outliers labeled on Figure~1 differ by more than 15-20\%, 
primarily because of the new spectral types and luminosity classes,  which can change the absolute magnitudes 
by $\Delta M_V = 0.3-0.8$. The agreement with 101 stars in Bowen \etal\ (2008) is not quite as good (Figure 2) 
because of differences in the adopted GOS photometry ($V$, $A_V$) and SpTs.   

\item A subset of 84 GOS stars provided updated stellar classifications and digital photometry, including 
extinction-corrected visual magnitudes requiring no assumptions about $R_V$.   Direct comparison of 
$D_{\rm GOS}$ and $D_{\rm Gaia}$ (Figures 4 and 5) illustrates the large dispersion between photometric 
and parallax distances for OB stars at  $d >1.5$~kpc.  Reliable {\it Gaia} parallax distances appear to require
parallax measurements with relative errors ($\sigma_{\varpi} / \varpi)$ less than 8\%.  

\item The {\it Gaia}-DR2 parallax distances track our photometric distances out to $d \approx 1.5$~kpc,
with increasing scatter at greater distances.  In the sub-sample of 81 GOS O-stars with reliable {\it Gaia}
parallaxes, $\sim30$\% (26 stars) have ratios $D_{\rm Gaia} / D_{\rm GOS}$ deviating by $\pm30$\% from 
unity.  This may reflect a broad, asymmetric distribution in parallax offsets about the applied standard value
of 0.03~mas, as well as possible errors in SpTs (and $M_V$).  With the current {\it Gaia}-DR2 parallax
uncertainties (Figure 5) we are unable to establish whether 0.03~mas is a better mean offset than the higher
value (0.05~mas) found for red giants.  Future data releases might allow us to make such distinctions.

\item Application of our new photometric-distance techniques to two OB associations resulted in 
revised estimates of their distances.   For nine O-type stars in Per OB1 we found 
$\langle D_{\rm GOS} \rangle = 2.99 \pm 0.14$~kpc and 
$\langle D_{\rm Gaia} \rangle = 2.47 \pm 0.57$~kpc.   
For 29 O-type stars in Carina OB1, we found 
$\langle D_{\rm GOS} \rangle = 2.60 \pm 0.28$~kpc and
$\langle D_{\rm Gaia} \rangle \approx 2.87 \pm 0.73$~kpc, 
with no statistical difference in distances to the four embedded star clusters 
(Tr~16, Tr~14, Tr~15, Coll~228).   

\item Our new photometric distances, $D_{\rm Shull}$ and $D_{\rm GOS}$, are based on a standard 
grid of absolute magnitudes from Bowen \etal\ (2008).  Changing to the grid of Martins \etal\ (2005),
with lower luminosities for O-type stars,  shifts these distance closer by $\sim 7$\%.  
For Car~OB1, we would then find $\langle D'_{\rm GOS} \rangle = 2.42 \pm 0.29$~kpc.

\end{enumerate}

\vspace{0.3cm}

From the dispersion of  distances shown in Figure 4, it  appears that the {\it Gaia} bright-star error distribution 
function may have a low-dispersion core and asymmetric high-dispersion wings.  Errors in the core correspond 
to the formal parallax errors, used for our quoted range in $D_{\rm Gaia}$.  For some stars with moderate or
large parallax errors $(\sigma_{\varpi} / \varpi > 8$\%) we see large differences between $D_{\rm Gaia}$ and 
$D_{\rm GOS}$.  To obtain distance agreement ($D_{\rm Gaia} \approx D_{\rm GOS}$) for these outliers, we 
would need to apply both positive and negative offsets (0.1--0.3~mas) from wings of the error distribution function.  
As noted above, 26 of 81 GOS stars in Figure~5 have $D_{\rm Gaia} / D_{\rm GOS}$ ratios deviating by 
$\pm 30$\% from unity.  The distribution appears to be asymmetric, with 20 outliers above the 1.3-ratio
 line, but only 6 stars below the 0.7-ratio line.  In determining distances to structures containing many O-type
 stars, one can sample the low-dispersion core of the error distribution and ignore the outliers, as we did for 
 Car~OB1 and Per~OB1.  Thus, the cluster membership issues suggested for Trumpler~16 (Davidson \etal\ 2018) 
 may be caused by broad wings in the parallax-offset distribution rather than cluster membership.

These new photometric distances toward OB-type stars will be adopted in our upcoming survey of \Htwo\ from 
\FUSE\ spectra.  The techniques developed in this pilot study can be extended to other O-stars in the GOS 
spectroscopic survey and compared to those used in previous \IUE, \FUSE, and \HST\ surveys of interstellar 
matter.  For selected OB associations, we can employ main-sequence fitting of  well-observed Galactic OB 
associations to determine a new $M_V$ calibration for OB stars.  We can use stars in Sco~OB2, and Ori~OB1 
as local distance anchors.   The OB stars in the LMC, with their precise distances and large pool of luminosity 
class I and III sources, will provide a crucial addition to the sample.  Previous studies, as well as recent
 theoretical stellar atmosphere models (Hainich \etal\ 2019)  have found almost no difference between the 
$M_V$-spectral type calibrations between LMC and Galactic OB stars.   This should allow us to assess the 
consistency of absolute magnitude tables (Vacca \etal\ 1996; Martins \etal\ 2005; Bowen \etal\ 2008).  
These consistency checks should be based on physical constraints, $L = 4 \pi R^2 \sigma T_{\rm eff}^4$ 
and $g = GM/R^2$ (with rotational corrections) for SpTs with SpT grids from O2 to B2 in all luminosity classes 
(V, IV, III, II, I).   The computed model atmospheres and flux grids may also contain errors comparable
to the systematic errors in {\it Gaia} parallax, arising from stellar rotation, outflows, MHD turbulence, and other
physical effects. 

The discussion in Appendix~A demonstrates that the resolution of outliers in spectroscopic distance comparisons 
could involve either changes in parallax offsets or modification of the GOS stellar classifications.  Tripp \etal\ (1993)
noted that some late O-type and early B-type stars could use UV line diagnostics  (Massa 1989) to distinguish the
correct luminosity classes.  Our new spectroscopic distances may benefit from such updates, in order to obtain better 
agreement between $D_{\rm phot}$ and $D_{\rm Gaia}$ after the next  {\it Gaia} data release (DR3) currently 
scheduled for late 2020.
  
\vspace{0.3cm}

\noindent
{\bf Acknowledgements.}  This work has made use of data from the European Space Agency (ESA) mission
 {\it Gaia}  (\url{https://www.cosmos.esa.int/gaia}), processed by the {\it Gaia} Data Processing and Analysis 
Consortium (DPAC, \url{https://www.cosmos.esa.int/web/gaia/dpac/consortium}). Funding for the DPAC
has been provided by national institutions, in particular the institutions participating in the {\it Gaia} 
Multilateral Agreement.   We thank Blair Savage, Ed Jenkins, Bill Vacca, Roberta Humphreys, Kris Davidson,
and Nathan Smith for helpful discussions on OB stars and  the Carinae Nebula.   We are also grateful to 
Jes\'us Ma\'iz Apell\'aniz for clarifications about the GOS spectroscopic survey and to Jeremy Darling and
Joel Zinn for advice about {\it Gaia} parallax uncertainties.  
 
\vspace{1.5cm}


\appendix
\section{Notes on Specific Stars }

Below, we provide brief discussion of photometric distance estimates for stars labeled as outliers 
on plots that compare our work  to previous photometric distances (Figure 1) or to {\it Gaia} 
parallax distances (Figures 4 and 5).  Data on {\it Gaia}-DR2 parallaxes ($\varpi \pm \sigma_{\varpi})$ 
are listed in Table 3.  We also elaborate on the assumptions made for complex binary systems and
uncertain classifications.  We are particularly interested in outliers in Figure~5 with distance ratios 
$D_{\rm Gaia} / D_{\rm GOS}$ deviating by $\pm30$\% from unity.  Are these the result of parallax 
errors or incorrect spectrophotometric distances?   How well are the SpTs classified?   How accurate 
are the absolute magnitudes?  

\subsection{\bf Outliers on Figure 1}  

\noindent
{\bf \#10 (HD 13268).}   Our photometric distance estimates, $D_{\rm Shull} = 2.77$~kpc and 
$D_{\rm GOS} = 2.80$~kpc, are larger than previous values, 1.75~kpc and 2.1~kpc from
Savage \etal\ (2001) and Bowen \etal\ (2008).   The difference arises from the GOS spectral 
type ON8.5~III (Sota \etal\ 2014) compared to ON8~V (Savage \etal\ 2001).  This produces a 
change in absolute magnitude $\Delta M_V = -0.80$ (ON8.5~III is more luminous).   
The {\it Gaia} distance is 1.61~kpc (range 1.50-1.74 kpc).  From the spectrum in Figure 14 of 
Sota \etal\ (2014), it is difficult to assess subtle differences between luminosity classes III and V 
in the diagnostic lines, \NIII\ (4634, 4640), \CIII\ (4650), \HeII\ (4686).  

\vspace{0.1cm}

\noindent
{\bf \#27 (HD 63005).}   Our photometric  distance estimate, $D_{\rm Shull} = 4.35$~kpc
(no GOS data), is lower than previous values, 5.23~kpc and 5.4~kpc from Savage \etal\ (2001) 
and Bowen \etal\ (2008).    We adopted a SpT of O7~V (Markova \etal\ 2011) compared to O6~V 
(Savage \etal\ 2001) who reference Garrison \etal\ (1977).   This produces $\Delta M_V = 0.30$ 
(O7~V is less luminous).  Markova \etal\ (2011) discuss a range in SpTs from O6.5~V to O7.5~V 
depending on resolution of the classifying spectrum;  we adopt O7~V as a median value.  

\vspace{0.1cm}

\noindent
{\bf \#31 (HD 69106).}  Our photometric distance estimates, $D_{\rm Shull} = 1.12$~kpc and 
$D_{\rm GOS} = 1.10$~kpc, are lower than previous values, 1.49 kpc and 1.5 kpc from 
Savage \etal\ (2001) and Bowen \etal\ (2008).   Our photometric distances are similar 
to $D_{\rm Gaia} = 1.24$~kpc (range 1.17--1.33~kpc).  We distrust the SpT of O9.7~II  
(Sota \etal\  2014), as it would imply $M_V = -5.83$ and a photometric distance $D = 2.9$~kpc, 
much greater than $D_{\rm GOS}$ and $D_{\rm Gaia}$.  Savage \etal\ (2001) used a SpT of 
B0.5~IV (Garrison \etal\ 1977) but we adopt B0.2~V (Markova \etal\ 2011) with $M_V = -3.70$.  
The SpT difference produces a change in absolute magnitude $\Delta M_V = 0.50$ (B0.2~V is 
less luminous than B0.5~IV).  The star is a fast rotator, which might influence the SpT.  

\vspace{0.1cm}

\noindent
{\bf \#34 (HD 74920).}  Our photometric distance estimates, $D_{\rm Shull} = 2.09$~kpc and 
$D_{\rm GOS} = 2.05$~kpc, differ from previous values, 1.50~kpc and 3.6~kpc from 
Savage \etal\ (2001) and Bowen \etal\ (2008).   Parallax gives $D_{\rm Gaia} = 2.65$~kpc 
(range 2.42--2.92 kpc).  
We adopt the GOS SpT of  O7.5~IV (Sota \etal\ 2014) rather than O8 (Savage \etal\ 2001) who 
reference Thackeray \& Andrews (1974) but quote no luminosity class.  Vijapurkar \& Drilling 
(1993) refer to this star as LSS~1148 and they assign it O7~III.   Differences in SpT are likely
responsible for  changes in absolute magnitude, particularly with no luminosity class assigned 
by either Savage \etal\ (2001) or Bowen \etal\ (2008).  

\vspace{0.1cm}

\noindent
{\bf \#35 (HD 89137).}  Our photometric distance estimates, $D_{\rm Shull} = 4.16$~kpc and 
$D_{\rm GOS} = 4.04$~kpc, are larger than previous values, 2.97~kpc and 3.1~kpc from
Savage \etal\ (2001) and Bowen \etal\ (2008).   The difference arises from the revised GOS 
spectral type ON9.7~II (Sota \etal\ 2014) compared to ON9.7~III (Savage \etal\ 2001) who 
reference Garrison \etal\ (1977).  This produces a change in absolute magnitude 
$\Delta M_V = -0.73$ (ON9.7~II is more luminous).   

\vspace{0.1cm}

\noindent
{\bf \#38 (HD 91651).}  Our photometric distance estimates, $D_{\rm Shull} = 4.31$~kpc and 
$D_{\rm GOS} = 3.98$~kpc, are larger than previous values, 2.81~kpc and 2.8~kpc from
Savage \etal\ (2001) and Bowen \etal\ (2008).   The difference arises from the revised GOS 
spectral type ON9.5~III (Sota \etal\ 2014) compared to O9~V (Savage \etal\ 2001) who 
reference Walborn (1973a).   This produces a change in absolute magnitude 
$\Delta M_V = -0.90$ (ON9.5~III is more luminous).   We also note a minor transcription
error in Savage \etal\ (2001), who quoted $V = 8.86$ rather than 8.84 (Schild \etal\ 1983).  

\vspace{0.1cm}

\noindent
{\bf \#43 (HD 93146A).}  Our photometric distance estimates, $D_{\rm Shull} = 2.84$~kpc and 
$D_{\rm GOS} = 2.30$~kpc, are smaller than previous values, 3.41~kpc and 3.5~kpc from 
Savage \etal\ (2001) and Bowen \etal\ (2008).   We adopt a SpT of  O7~V (Sota \etal\ 2014) 
whereas Savage \etal\ (2001) list O6.5~V.  This produces a change in absolute magnitude 
$\Delta M_V = 0.15$ (O7~V is less luminous). This  is a binary system, with HD~93146A (O7~V) 
and HD~93146B (O9.7~IV).  A member of Car~OB1 and the Coll~228 cluster, with anomalous 
reddening ($R_V \approx 4$), this star is likely to lie at 2.3-2.6~kpc (see Section~4.1 and Table 4).  

\vspace{0.1cm}

\noindent
{\bf \#49 (HD 93843).}  Our photometric distance estimates, $D_{\rm Shull} = 3.21$~kpc and 
$D_{\rm GOS} = 2.85$~kpc, are smaller than previous values, 3.68~kpc and 3.5~kpc from Savage 
\etal\ (2001) and Bowen \etal\ (2008).    Parallax gives $D_{\rm Gaia} = 2.43$~kpc (range 
2.24--2.66 kpc).   We adopt a SpT of O5~III (Sota \etal\ 2014), the same as in Savage \etal\ (2001).  
The difference in distances does not arise from the adopted photometry.  We adopt $V = 7.30$ 
(star \#472 in Schild \etal\ 1983) and $E(B-V) = 0.28$.  Savage \etal\ (2001) list $V = 7.34$ from 
the same reference (likely mis-transcribed).  GOS photometry (Ma\'iz Apell\'aniz \& Barb\'a 2018)
give $V_J = 7.32$ and $A_V = 1.146$.  This sight line has  anomalous reddening with 
$A_V / E(B-V) = 4.1$, which accounts for our lower distance $D_{\rm GOS} = 2.85$~kpc.  

\vspace{0.1cm}

\noindent
{\bf \#68 (HD 115071).} Our photometric distance estimates, $D_{\rm Shull} = 2.05$~kpc and 
$D_{\rm GOS} = 1.87$~kpc, are much larger than previous values, 1.01~kpc and 1.2~kpc from
Savage \etal\ (2001) and Bowen \etal\ (2008).   The difference arises from the revised GOS 
spectral type O9.5~III (Sota \etal\ 2014) compared to B0.5~V (Savage \etal\ 2001) who 
reference Garrison \etal\ (1977).   This produces a large change in absolute magnitude 
$\Delta M_V = -1.65$ (O9.5~III is more luminous).  We note that $D_{\rm Gaia} = 1.98$~kpc 
(range 1.84--2.14~kpc) is consistent with our photometric distances.  

\vspace{0.1cm}

\noindent
{\bf \#73 (HD 124979).}  Our photometric distance estimates, $D_{\rm Shull} = 3.05$~kpc and 
$D_{\rm GOS} = 3.09$~kpc, are larger than previous values, 2.18~kpc and 2.8~kpc from Savage 
\etal\ (2001) and Bowen \etal\ (2008).   We assume a SpT of O7.5~IV (Sota \etal\ 2014) compared to 
O8~V (Savage \etal\ 2001) who reference Hill \etal\ (1974).  This produces a change in absolute magnitude
 $\Delta M_V = -0.56$ (O7.5~IV is more luminous than O8~V).   

\vspace{0.1cm}

\noindent
{\bf \#89 (HD 164816).}  Our distance estimates, $D_{\rm Shull} = 1.13$~kpc and 
$D_{\rm GOS} = 1.08$~kpc, are smaller than the value, 1.69~kpc, in Savage \etal\ (2001).  
They are comparable to $D_{\rm Gaia} = 1.14$~kpc (range 1.06--1.24~kpc) based on parallax 
$0.8442 \pm 0.0704$~mas (8.3\% formal error) and 0.03 mas offset.  We adopt a SpT of O9.5~V
(Sota \etal\ 2014) with $M_V = -4.15$ compared to O9.5~III-IV (Savage \etal\ 2001) with $M_V = -4.94$.   
This produces a change in absolute magnitude $\Delta M_V = +0.79$ (O9.5~V is less luminous).   
The difference hinges on the star's luminosity class (V vs.\  III-IV).  

\vspace{0.1cm}

\noindent
{\bf \#99 (HD 168941).}  Our photometric distance estimates, $D_{\rm Shull} = 3.75$~kpc and 
$D_{\rm GOS} = 3.72$~kpc, are smaller than previous values, 5.79~kpc and 6.8~kpc from
Savage \etal\ (2001) and Bowen \etal\ (2008).   The difference arises from the revised GOS 
spectral type O9.5~IV (Sota \etal\ 2014) with $M_V = -4.68$ compared to O9.5~II-III  (Savage 
\etal\ 2001; Tripp \etal\ 1993) with $M_V \approx -5.52$.  This produces a change in absolute magnitude
$\Delta M_V = +0.84$ (O9.5~IV is less luminous). Tripp \etal\ (1993) used UV lines (\SiII, \SiIII, \SiIV, \CIV, 
 \NIV) as classification diagnostics (Massa 1989) and quoted a distance of 6.1~kpc.   {\it Gaia}-DR2 gives 
 $D_{\rm Gaia} = 2.32$~kpc range 1.96--2.84 kpc) based on parallax $0.4018 \pm 0.0792$ (20\% formal 
 error).   The large difference in spectroscopic distances hinges on the star's luminosity class (IV vs.\  II-III).  


\subsection{\bf Outliers on Figure 4}  

\noindent
{\bf \#3 (CPD 59$^{\circ}$2600).}   Our photometric distance estimates, $D_{\rm Shull} = 2.71$~kpc
and $D_{\rm GOS} = 2.11$~kpc, are lower than previous values, 2.84~kpc and 2.9~kpc from 
Savage \etal\ (2001) and Bowen \etal\ (2008).   The GOS spectral type O6~Vf (Sota \etal\ 2014) is the same 
as used in previous studies.  Our lower value $D_{\rm GOS} = 2.11$~kpc is a result of anomalous 
extinction in Carina.  A member of Car~OB1 and the Tr~16 cluster with $R_V \approx 4$, this star probably 
lies at 2.3--2.6~kpc (see Section~4.1 and Table 4).  Our photometric distances are much smaller than 
$D_{\rm Gaia} = 5.11$~kpc (range 4.28--6.36~kpc) based on parallax $0.1655 \pm 0.0382$~mas 
(23\% formal error) and 0.03 mas offset.  Increasing the parallax offset to 0.20-0.25 mas would bring the 
parallax distance into better agreement with the photometric distance of Tr~16 and the one-to-one line
($D_{\rm Gaia} = D_{\rm GOS}$).  

\vspace{0.1cm}

\noindent
{\bf \#10 (HD 13268).}   Our photometric distance estimates, $D_{\rm Shull} = 2.77$~kpc and 
$D_{\rm GOS} = 2.80$~kpc, are larger than $D_{\rm Gaia} = 1.61$~kpc (range 1.50--1.74~kpc) based 
on parallax $0.5906 \pm 0.0466$~mas (7.9\% formal error) and  0.03 mas offset.  Increasing the parallax
offset to larger values would give worse agreement,  as star \#10 lies below the one-to-one line.  
This may be a case for a negative parallax offset ($-0.23$~mas).  Alternatively, the SpT (ON~8.5~III) may 
be incorrect.  

\vspace{0.1cm}

\noindent
{\bf \#11 (HD 13745).}   Our photometric distance estimates, $D_{\rm Shull} = 2.80$~kpc and 
$D_{\rm GOS} = 2.81$~kpc, are larger than $D_{\rm Gaia} = 2.13$~kpc (range 1.91--2.39~kpc) 
based on parallax $0.4405 \pm 0.0528$~mas (12.0\% formal error) and 0.03 mas offset.  Increasing 
the parallax offset would give worse agreement, as star \#11 lies below the one-to-one line.  This may
suggest a negative parallax offset ($-0.08$~mas).  

\vspace{0.1cm}

\noindent
{\bf \#12 (HD 14434).}   Our photometric distance estimates, $D_{\rm Shull} = 2.95$~kpc and 
$D_{\rm GOS} = 2.98$~kpc, are somewhat larger than $D_{\rm Gaia} = 2.37$~kpc 
(range 2.13--2.67~kpc) based on parallax $0.3912 \pm 0.0473$~mas (12.1\% formal error) and 
0.03 mas offset.  Increasing the parallax offset to larger values would give worse agreement, 
as star \#12 lies below the one-to-one line.  This may be a case for a negative parallax offset
($-0.05$~mas).  

\vspace{0.1cm}

\noindent
{\bf \#18 (HD 41161).}  Our photometric distances, $D_{\rm Shull} = 1.35$~kpc and 
$D_{\rm GOS} = 1.16$~kpc, as well as $D_{\rm Savage} = 1.23$~kpc are lower than 
$D_{\rm Gaia} = 1.52$~kpc (range 1.40--1.66~kpc) based on parallax 
$0.6284 \pm 0.0575$~mas (9.2\% formal error) with 0.03 mas  offset.  Increasing the parallax 
offset to 0.20~mas would bring the star into closer agreement with the one-to-one line. 

\vspace{0.1cm}

\noindent
{\bf \#32 (HD 73882).}  Our photometric distance estimates, $D_{\rm Shull} = 1.00$~kpc and 
$D_{\rm GOS} = 0.83$~kpc, are similar to 0.79~kpc (Savage \etal\ 2001) but much larger than 
$D_{\rm Gaia} = 0.34$~kpc (range 0.30--0.41~kpc) based on parallax $2.8856\pm0.4507$~mas 
(16\% formal error) with 0.03 mas  offset.  Our SpT of O8.5~IV from Sota \etal\ (2014) differs from
O8~V in Savage et al. (2001) who quote Garrison \etal\ (1977).  This produces a change in absolute 
magnitude $\Delta M_V = -0.32$ (O8.5~IV is more luminous).  GOS photometry gives $V_J = 7.25$ 
(Ma\'iz Apell\'aniz \& Barb\'a 2018) similar to $V = 7.22$ (Savage \etal\ 2001) who quote 
Schild \etal\ (1983).  Our derived value $E(B-V ) = 0.69$ agrees with Savage \etal\ (2001).  
The system is listed as an eclipsing binary, which could affect  parallax measurements.

\vspace{0.1cm}

\noindent
{\bf \#34 (HD 74920).}  Our photometric distance estimates, $D_{\rm Shull} = 2.09$~kpc and 
$D_{\rm GOS} = 2.05$~kpc, are smaller than $D_{\rm Gaia} = 2.65$~kpc (range 2.42--2.92 kpc) 
based on parallax $0.3479 \pm 0.0358$~mas (10\% formal error) and 0.03 mas  offset.  Increasing the 
parallax offset to 0.14~mas would bring the star into closer agreement with the one-to-one line. 
As noted in Section 6.1, differences in SpT (O7.5~IV vs.\ O7~III) could produce a sizeable change 
in absolute magnitude.

\vspace{0.1cm}

\noindent
{\bf \#38 (HD 91651).}  Our photometric distance estimates, $D_{\rm Shull} = 4.31$~kpc and 
$D_{\rm GOS} = 3.98$~kpc, are larger than $D_{\rm Gaia} = 1.83$~kpc (range 1.69--2.00 kpc)
based on parallax $0.5167 \pm 0.0455$~mas (8.8\% formal error) and 0.03 mas offset.  Increasing
the parallax offset would give worse agreement, as star \#38 lies below the one-to-one line.  This 
may be a case for a negative parallax offset ($-0.27$~mas) or an incorrect SpT (see Section 6.1).  

\vspace{0.1cm}

\noindent
{\bf \#43 (HD 93146A).}  Our photometric distance estimates, $D_{\rm Shull} = 2.84$~kpc and 
$D_{\rm GOS} = 2.30$~kpc, are smaller than $D_{\rm Gaia} = 3.14$~kpc (range 2.80--3.58~kpc)
based on parallax $0.2880 \pm 0.0386$~mas (13.4\% formal error) and 0.03 mas offset. 
A member of Car~OB1 and the Coll 228 cluster with anomalous reddening ($R_V \approx 4$), 
this star is likely to lie at 2.3-2.6~kpc (see Section~4.1 and Table~4).  Increasing the parallax 
offset to 0.10~mas would bring the star into better agreement with the cluster distance and the 
one-to-one line. 

\vspace{0.1cm}
 
\noindent
{\bf \#50 (HD 96670).}  We find good agreement in all three  photometric distances,
$D_{\rm Shull} = 2.78$~kpc, $D_{\rm GOS} = 2.67$~kpc, and $D_{\rm Savage} = 2.88$~kpc, 
all smaller than $D_{\rm Gaia} = 3.59$~kpc (range 3.18--4.09~kpc) based on parallax 
$0.2418 \pm 0.0344$ (14\% formal error) and 0.03 mas offset.  Increasing the parallax offset 
to 0.12~mas would bring the star into better agreement with the one-to-one line.   This star
did not appear with a SpT in the GOS papers (Sota \etal\ 2011, 2014).  We adopted O8~Ibf 
(Garrison \etal\ 1977) with GOS photometry from Ma\'iz Apell\'aniz \& Barb\'a (2018).  

\vspace{0.1cm}

 \noindent
{\bf \#52 (HD 96917).}  We find good agreement in all three  photometric distances,
$D_{\rm Shull} = 2.61$~kpc, $D_{\rm GOS} = 2.46$~kpc, and $D_{\rm Savage} = 2.68$~kpc, 
all smaller than $D_{\rm Gaia} = 4.41$~kpc (range 3.72--5.43~kpc) based on parallax 
$0.1966 \pm 0.0423$ (22\% formal error) and 0.03 mas offset.  Increasing the parallax offset 
to 0.20~mas would bring the star into agreement with the one-to-one line. 

\vspace{0.1cm}

\noindent
{\bf \#60 (HD 101131).}   Our photometric distance estimates, $D_{\rm Shull} = 2.03$~kpc and 
$D_{\rm GOS} = 1.78$~kpc, are similar to previous distances, $D_{\rm Savage} = 1.91$~kpc and 
$D_{\rm Bowen} = 2.0$~kpc.  All are smaller than $D_{\rm Gaia} = 2.44$~kpc (range 2.07--2.97~kpc) 
based on parallax $0.3795 \pm 0.0729$~mas (19.2\% formal error) and 0.03 mas offset.  This star 
has anomalous reddening ($R_V \approx 4$) based on $A_V = 1.228$ from GOS (Ma\'iz Apell\'aniz 
\& Barb\'a 2018) and our value $E(B-V) = 0.31$.  Increasing the parallax 
offset to 0.15~mas would bring the star into better agreement with the one-to-one line. 

\vspace{0.1cm}

\noindent
{\bf \#61 (HD 101190).} Our photometric distance estimates, $D_{\rm Shull} = 2.16$~kpc and 
$D_{\rm GOS} = 2.09$~kpc, are similar to $D_{\rm Savage} = 1.95$~kpc and $D_{\rm Bowen} = 2.1$~kpc, 
but considerably smaller than $D_{\rm Gaia} = 3.06$~kpc (range 2.73--3.47~kpc) based on parallax 
$0.2971 \pm 0.0389$~mas (13.1\% formal error) and 0.03 mas offset.   Increasing the parallax offset to 
0.20~mas would bring the star into better agreement with the one-to-one line. The composite spectrum 
indicates a B-star companion that could affect the parallax measurement.  

\vspace{0.1cm}

\noindent
{\bf \#62 (HD 101205).} Our photometric distance estimates, $D_{\rm Shull} = 1.68$~kpc and 
$D_{\rm GOS} = 1.64$~kpc, are similar to previous distances, $D_{\rm Savage} = 1.47$~kpc and 
$D_{\rm Bowen} = 2.4$~kpc, but smaller than $D_{\rm Gaia} = 2.63$~kpc (range 1.94--4.08~kpc) 
based on parallax $0.3557 \pm 0.1354$~mas (39\% formal error) and 0.03 mas offset.  Increasing
 the parallax offset to 0.25~mas would bring the star into better agreement with the one-to-one line.

\vspace{0.1cm}

\noindent
{\bf \#64 (HD 101413).}  Our photometric distance estimates, $D_{\rm Shull} = 2.33$~kpc and 
$D_{\rm GOS} = 2.13$~kpc, are similar to $D_{\rm Savage} = 2.12$~kpc and 
$D_{\rm Bowen} = 2.4$~kpc.  They are somewhat larger than $D_{\rm Gaia} = 1.78$~kpc 
(range 1.65--1.94~kpc) based on parallax $0.5304 \pm 0.0440$~mas (8.3\% formal error) and 
0.03 mas offset.  Increasing the parallax offset to larger values would give worse agreement, 
as star \#64 lies below the one-to-one line.  This may be a case for a negative parallax offset
($-0.06$~mas).  

\vspace{0.1cm}

\noindent
{\bf \#65 (HD 101436).}  Our photometric distance estimates, $D_{\rm Shull} = 1.93$~kpc and 
$D_{\rm GOS} = 1.85$~kpc, are similar to $D_{\rm Savage} = 2.17$~kpc and 
$D_{\rm Bowen} = 2.2$~kpc, but smaller than $D_{\rm Gaia} = 2.81$~kpc (range 2.18--3.95~kpc) 
based on parallax $0.3257 \pm 0.1027$~mas (32\% formal error) and 0.03 mas offset.   Increasing the 
parallax offset to 0.20~mas would bring the star into better agreement with the one-to-one line. 

\vspace{0.1cm}

\noindent
{\bf \#72 (HD 124314A).} Our photometric distance estimates, $D_{\rm Shull} = 1.27$~kpc and 
$D_{\rm GOS} = 1.25$~kpc, are similar to $D_{\rm Savage} = 1.15$~kpc and slightly lower than
$D_{\rm Bowen} = 1.4$~kpc (range 1.2--1.7~kpc).  Our SpT of O6~IV (Sota \etal\ 2014) is similar 
to the O6~V adopted by Savage \etal\ (2001) quoting Walbrn (1973a).  The photometric distances 
are smaller than $D_{\rm Gaia} = 1.72$~kpc (range 1.62--1.83~kpc) based on parallax 
$0.5530 \pm 0.0360$~mas (6.5\% formal error) and 0.03 mas offset.   Increasing the parallax offset 
to 0.25~mas would bring the star into better agreement with the one-to-one line.   Because this is a 
multiple-star system (Sota \etal\ 2014) with HD~124314A separated by $2.5"$ from a close binary
HD~124314BaBb classified as O9.2~IV, the parallax measurements could be affected.   

\vspace{0.1cm}

\noindent
{\bf \#86 (HD 161807).} Our photometric distances, $D_{\rm Shull} = 1.86$~kpc and 
$D_{\rm GOS} = 1.94$~kpc, are similar to $D_{\rm Bowen} = 2.1$~kpc.  This star was not studied 
by Savage \etal\ (2001).  Our adopted SpT of O9.7~III (Sota \etal\ 2014) is slightly earlier than 
the B0~III in Bowen \etal\ (2008) quoting Garrison \etal\ (1977).  The photometric distances are larger 
than $D_{\rm Gaia} = 1.27$~kpc (range 1.18--1.37~kpc) based on parallax $0.7576 \pm 0.0602$~mas 
(8.0\% formal error) and 0.03 mas offset.   Increasing the parallax offset to larger values would give 
worse agreement, as star \#86 lies below the one-to-one line.  This may suggest a negative parallax 
offset ($-0.23$~mas).   This star is listed as an eclipsing binary (Garrison \etal\ 1983).  

\vspace{0.1cm}

\noindent
{\bf \#91 (HD 165246).} Our photometric distances are $D_{\rm Shull} = 1.64$~kpc and 
$D_{\rm GOS} = 1.38$~kpc.  This star was not studied by either Savage \etal\ (2001) or 
Bowen \etal\ (2008).  Our adopted SpT of O8~V (Sota \etal\ 2014) is the same as that of 
Garrison \etal\ (1977).  The photometric distances are smaller than $D_{\rm Gaia} = 1.88$~kpc 
(range 1.55--2.39~kpc) based on parallax $0.5011 \pm 0.1131$~mas (23\% error) 
and 0.03 mas offset.   Increasing the parallax offset to 0.22~mas would bring the star into better 
agreement with the one-to-one line. 

\vspace{0.1cm}

\noindent
{\bf \#96 (HD 167771).} Our photometric distances, $D_{\rm Shull} = 1.50$~kpc and 
$D_{\rm GOS} = 1.40$~kpc are similar to the value 1.43~kpc (Savage \etal\ (2001).  Our adopted 
SpT of  O7~III (Sota \etal\ 2014) is the same as that in Savage \etal\ (2001) quoting Walborn (1972). 
The photometric distances are smaller than $D_{\rm Gaia} = 1.88$~kpc (range 1.55--2.39~kpc) 
based on parallax $0.5202 \pm 0.0498$~mas (9.6\% error) and 0.03 mas offset.   Increasing 
the parallax offset to 0.19~mas would bring the star into better agreement with the one-to-one line. 

\vspace{0.1cm}

\noindent
{\bf \#97 (HD 167971).} Our photometric distances are $D_{\rm Shull} = 1.68$~kpc and 
$D_{\rm GOS} = 1.42$~kpc.  This star was not studied by either Savage \etal\ (2001) or 
Bowen \etal\ (2008).  Our adopted SpT of O8~Ia (Sota \etal\ 2011) is comparable to previous
values of O8~f  (Hiltner 1956) and O8~Ib (Walborn 1972).  The photometric distances are 
smaller than $D_{\rm Gaia} = 1.92$~kpc (range 1.58--2.43~kpc) based on parallax 
$0.4918 \pm 0.1106$~mas (22.5\% error) and 0.03 mas offset.   Increasing the parallax offset 
to 0.21~mas would bring the star into better agreement with the one-to-one line. 

\vspace{0.1cm}

\noindent
{\bf \#99 (HD 168941).}  Our photometric distance estimates, $D_{\rm Shull} = 3.75$~kpc and 
$D_{\rm GOS} = 3.72$~kpc, are smaller than previous values, 5.79~kpc and 6.8~kpc from
Savage \etal\ (2001) and Bowen \etal\ (2008), owing to an updated SpT (see Section 6.1 above).  
They are larger than  $D_{\rm Gaia} = 2.32$~kpc  (range 1.96--2.84 kpc) based on parallax 
$0.4018 \pm 0.0792$~mas (20\% formal error).   Increasing the parallax offset to larger values would 
give worse agreement, as star \#99 lies below the one-to-one line.   This may be a case for
a negative parallax offset ($-0.13$~mas). 

\vspace{0.1cm}

\noindent
{\bf \#120 (HD 206267).}  Our photometric distance estimates, $D_{\rm Shull} = 0.62$~kpc and 
$D_{\rm GOS} = 0.73$~kpc, are similar to $D_{\rm Savage} = 0.72$~kpc, but smaller than 
$D_{\rm Gaia} = 1.08$~kpc (range 0.86--1.45~kpc) based on parallax $0.8952 \pm 0.2356$~mas
(26\% formal error) and 0.03 mas offset.   A very large (0.40 mas) parallax offset would be 
required to bring the star into agreement with the one-to-one line.  This star was not classified in 
the GOS papers (Sota \etal\ 2011, 2014).  We adopt a SpT of O6~V from Saurin \etal\ (2012) who 
associate this star with \HII\ region IC~1396 and the embedded cluster Trumpler~37 at mean 
distance $d = 800 \pm 60$~pc.    Pan \etal\ (2004) place this star in Cep~OB2 and list $V = 5.62$, 
$E(B-V) = 0.51$, and $d = 750$~pc.  Our photometric distance estimates (0.62--0.73~pc) agree 
with these observations.  The GOS photometry (Ma\'iz Apell\'aniz \& Barb\'a 2018) listed $
V_J = 5.688$ and $A_V = 1.584$, so that  $R_V \approx 3.1$.  This star is a spectroscopic binary 
(O6~V + O9~V) which might affect  parallax measurements.  

\vspace{0.1cm}

\noindent
{\bf \#125 (HD 209339).}  Our photometric distance estimates, $D_{\rm Shull} = 1.18$~kpc and 
$D_{\rm GOS} = 1.08$~kpc, were based on a SpT of O9.7~IV (Sota \etal\ 2014).  This star was not 
studied by Savage \etal\ (2001) or Bowen \etal\ (2008).  Our photometric distances are larger than 
$D_{\rm Gaia} = 0.82$~kpc (range 0.80--0.85~kpc) based on parallax $1.1836 \pm 0.0303$~mas 
(2.6\% formal error) and 0.03 mas offset.   Increasing the parallax offset to larger values would give 
worse agreement, as star \#125 lies below the one-to-one line.  This may be a case for a negative 
parallax offset ($-0.25$~mas).  

\vspace{0.1cm}

\noindent
{\bf \#127 (HD 210839).}  The O-supergiant $\lambda$~Cep has photometric distance estimates, 
$D_{\rm Shull} = 0.81$~kpc and $D_{\rm GOS} = 0.87$~kpc,  based on a SpT of O6.5~I(n)fp
(Sota \etal\ 2011) with photometry $V = 5.05$ and $B-V = 0.25$ (Hiiltner 1956) in agreement 
with the GOS digital photometry (Ma\'iz Apell\'aniz \& Barb\'a 2018).  We assume intrinsic color
$(B-V)_0 = -0.32$ and $E(B-V) = 0.57$, and we adopt a luminosity class Ib ($M_V = -6.25$).
Previous studies found distances of 0.81~kpc (Savage \etal\ 2001) and 1.1~kpc (Bowen \etal\ 2008) 
using a slightly earlier SpT of O6~Inf (Walborn 1973a).  Bouret \etal\ (2012) classified $\lambda$~Cep 
as O6~I(n)fp and modeled the stellar parameters with $V = 5.05$, $B-V = 0.192$, $E(B-V) = 0.513$, 
$M_V = -6.43$, and $d = 0.95\pm0.10$~kpc.  Pan \etal\ (2004) adopted O6~Iab with $V = 5.09$, 
$E(B-V) = 0.56$, and $d = 800$~pc for the Cep~OB2 association.  Gvaramadze \& Gualandris (2011)
suggest that $\lambda$~Cep is a runaway star expelled from Cep~OB3.  Distances estimates to 
Cep~OB3 range from 725~pc (Blauuw \etal\ 1959) to 870~pc (Humphreys 1978).  A detailed study 
(Simonson 1968) places $\lambda$~Cep in the Cep~OB2 association, whose distance was 
estimated at $d = 615 \pm 35$~pc (de~Zeeuw \etal\ 1999) although possibly as large as 800~pc.
A SpT of O6~Iab would correspond to $M_V = -6.60$, slightly more luminous than 
$M_V = -6.43^{+0.11}_{-0.12}$ in the model of Bouret \etal\ (2012).  These photometric distances 
are all larger than $D_{\rm Gaia} = 0.61$~kpc (range 0.56--0.66~kpc) based on parallax 
$1.6199 \pm 0.1265$~mas (7.8\% formal error) and 0.03 mas offset.   The {\it Hipparcos} parallax 
is similar, $1.65 \pm 0.22$~mas (van Leeuwen 2007).  Because star \#127 lies below the one-to-one 
ratio line (Figures~4 and 5), ringing $D_{\rm Gaia}$ into agreement with photometric distances
would require a large negative parallax offset ($-0.5$~mas).   Quite possibly, the SpT of O6.5~I 
and the corresponding $M_V$ need to be re-examined. 

\vspace{0.1cm}

\noindent
{\bf \#129 (HD 216532).}  Our photometric distance estimates, $D_{\rm Shull} = 0.918$~kpc and 
$D_{\rm GOS} = 0.944$~kpc, are larger than $D_{\rm Gaia} = 0.734$~kpc (range 0.719--0.751~kpc)
based on parallax $1.3316 \pm 0.0299$~mas (2.3\% formal error) and 0.03 mas offset.  Increasing 
the parallax offset to larger values would give worse agreement, as star \#129 lies below the 
one-to-one line.  This may be a case for a negative parallax offset ($-0.25$~mas), even though the 
formal parallax errors are small.  Reconciling the distances might require shifting the star to a later 
SpT than O8.5~V (Sota \etal\ 2011) to produce a fainter absolute magnitude.  
However, Hiltner (1956), Morgan \etal\ (1955), and Garrison (1970) all list this star as O8~V, which 
would worsen the distance discrepancy.  Our adopted $E(B-V) = 0.85$ appears normal, with 
$R_V = 3.03$ based on $A_V = 2.576$ (Ma\'iz Apell\'aniz \& Barb\'a 2018).  

\vspace{0.1cm}

\noindent
{\bf  \#133 (HD~218915),}  We find excellent agreement in photometric distances, with 
$D_{\rm Shull} = 3.66$~kpc, $D_{\rm GOS} = 3.62$~kpc, and $D_{\rm Savage} = 3.63$~kpc, 
The {\it Gaia} distance of 5.6~kpc  (range 5.0--9.4~kpc)  is based on parallax 
$0.1241 \pm 0.0472$~mas (38\% formal error) and 0.03~mas offset.   Increasing the offset 
to 0.15~mas would bring the star into better agreement with the one-to-one line. 

\vspace{0.1cm}

\noindent
{\bf \#138 (HD~308813).}  We find agreement in our new photometric distances, 
$D_{\rm Shull} = 2.93$~kpc and $D_{\rm GOS} = 2.66$~kpc, and  previous estimates of
2.92~kpc (Savage \etal\ 2001) and 3.1~kpc (Bowen \etal\ 2008).  The larger {\it Gaia} distance 
of 4.56~kpc (range 3.97--5.53~kpc) is based on parallax $0.1894 \pm 0.0386$~mas (20\% formal 
error) and 0.03~mas offset.  Increasing the parallax offset to 0.10~mas would
bring the star into better agreement with the one-to-one line.


\subsection{\bf  Other stars with SpT discrepancies}

\noindent
{\bf \#8 (HD~5005A).}   HD~5005 is a quadruple system (A,B,C,D, all O-type stars).  Our photometric 
distances for HD~5005A, $D_{\rm Shull} = 3.13$~kpc and $D_{\rm GOS} = 2.89$~kpc, are 
larger than 2.28~kpc (Savage \etal\ 2001) and comparable to 2.9~kpc (Bowen \etal\ 2008).  
The difference arises from the GOS spectral type O4~V (Sota \etal\ 2011) compared to O6.5~V 
(Savage \etal\ 2001) who reference Walborn (1973a).  This produces a change in absolute magnitude
$\Delta M_V = -0.60$ (O4~V is more luminous).   The GOS distance of 2.89~kpc is consistent with the 
water-maser distance of $2.82^{+0.26}_{-0.22}$~kpc (Choi \etal\ 2014) to the star-forming region 
G123.06-6.50 in the adjoining nebula NGC~281.  {\it Gaia}-DR2 also provides a consistent 
(offset-corrected) distance $D_{\rm Gaia} = 2.71$~kpc.

\vspace{0.1cm}

\noindent
{\bf \#42 (HD 93129A).}   This star was classified by Walborn (1982) as O3If* and later as the
prototypical O2~If$^*$ star (Walborn \etal\ 2002).  This is the most massive star in the core of 
Trumpler~14, separated by $2.5''$ from HD~93129B (O3.5~V).  We analyze HD~93129A, which 
was resolved into binary components classified by Sota \etal\ (2014) as O2~If$^*$ (HD~93129Aa) 
and O3~IIIf$^*$ (HD~93129Ab).  The GOS photometry (Ma\'iz Apell\'aniz \& Barb\'a 2018) 
includes all 3 stars (Aa, Ab and B) with $V_{J,0} = 4.825$ and $A_V = 2.199$ (see our Table 4).  
Gruner \etal\ (2019) combined observations (HST and VLT) with theoretical spectral decomposition
to produce separate parameters for each component:  
HD 93129Aa [$V = 7.65$, $M_V = -6.09$, $E(B-V) = 0.57$] and 
HD 93129Ab [$V = 8.55$, $M_V = -5.21$, $E(B-V) = 0.57$].  These result in consistent 
photometric distances, $D_{\rm phot} = 2.48$~kpc (Aa) and 2.50~kpc (Ab), for an extinction law 
with $R_V = 3.1$.  If we adopted the anomalous extinction ($R_V = 4.0$) proposed for the Carina 
Nebula, these distances would drop to 1.96~kpc and 1.98~kpc, respectively.  In the combined 
AB-system, these data imply photometric distances
 $D_{\rm GOS} = 2.21$~kpc for $M_V = -6.90$ (Bowen \etal\ 2008) and
 $D'_{\rm GOS} = 1.72$~kpc for $M_V = -6.35$  (Martins \etal\ 2005).

\vspace{0.1cm}

\noindent
{\bf \#46 (HD 93206A).}   The multiple-star system QZ Car (HD 93206AB) is the brightest object in the 
Collinder~228 star cluster in the southern part of the Carina Nebula and a double (SB1+SB1)  binary
(Parkin \etal\ 2011), consisting of system~A (O9.7~Ib + b2 v) and system B (O8~III  + o9 v).  
The lower-case letters indicate that 
these are not true spectral classifications.  The GOS spectroscopic survey gives a combined classification 
of O9.7~Ib (Sota \etal\ 2014) with $V_{J,0} = 4.206$ and $A_V = 2.106$ (Ma\'iz Apell\'aniz \& Barb\'a 2018) 
presumably including all 4 stars in HD 93206AB.    Our distance estimates, $D_{\rm Shull} = 2.23$~kpc and 
$D_{\rm GOS} = 1.48$~kpc, are based on the Aa component (O9.7Ib) with $V = 6.31$ and $M_V = -6.18$, 
corrected for the luminosity ratio, $L_2/L_1 = 0.535$ of the two brightest stars (O8~III and O9.5~Ib).  This 
increases the distance by a factor $[1+(L_2/L_1)]^{1/2} = 1.24$.  Our distances are similar to the value 
(1.78~kpc) from Savage \etal\ (2001) but lower than the distance (2.3--2.6~kpc) expected if this system lies
 in the Carina Nebula (see Table 4).  

\vspace{0.1cm}

\noindent
{\bf \#48 (HD 93250).}  Our distance estimates, $D_{\rm Shull} = 2.62$~kpc and $D_{\rm GOS} = 2.20$~kpc, 
are comparable to previous values, 1.89~kpc and 2.3~kpc from Savage \etal\ (2001) and Bowen \etal\ (2008).   
Some difference arises from the revised GOS spectral types, O4~III (Sota \etal\ 2014) or O4~IV (Maiz Apellaniz 
\etal\ 2016), compared to O3~V in Savage \etal\ (2001).   Our photometric distances are based on O4~III.
They would change if we adopted the anomalous reddening in Carina with $R_V = 4.0$ or selected the 
Martins \etal\ (2005) grid of $M_V$ (see discussion in Section 4.1).  

\vspace{0.1cm}

\noindent
{\bf \#82 (HD 154368).}  Our distance estimates, $D_{\rm Shull} = 1.08$~kpc and $D_{\rm GOS} = 1.08$~kpc, 
are smaller than the value, 1.35~kpc, in Savage \etal\ (2001).  We adopt a SpT of O9.2~Iab (Sota \etal\ 2014) 
with $M_V = -6.546$ compared to O9~Ia (Savage \etal\ 2001) with $M_V = -7.00$.   This produces a change 
in absolute magnitude $\Delta M_V = 0.454$ (O9.2~Iab is less luminous).   The O9~Ia type comes from
Hiltner \etal\ (1969), whereas Garrison \etal\ (1977) give O9.5~Iab, which has $M_V = -6.54$. 

\vspace{0.1cm}

\noindent
{\bf \#105 (HD 179406).}  This B-type star, also known as 20~Aql, was not in the surveys of Savage \etal\ (2001) 
or Bowen \etal\ (2008).  There has been some controversy in the literature over its SpT, with luminosity classes 
ranging from V to II.    Lesh (1968) classifies it as B3~V, while Braganca \etal\  (2012) classified it as B2/3~II, and 
Buscombe (1962) listed B3~IV.  We adopt the latter classification of B3~IV ($M_V = -2.30$) with $B = 5.46$ and 
$V = 5.34$ from Braganca \etal\  (2012).   From $(B-V)_0 = -0.20$, we then derive $E(B-V) = 0.33$ and a 
photometric distance $D_{\rm Shull} = 211$~pc rather than 147 pc for B3~V ($M_V = -1.52$).  
Both are smaller than $D_{\rm Gaia} = 280$~pc (range 267-295~pc) based on parallax $3.5374 \pm 0.1720$~mas
(4.9\% formal error) and 0.03~mas offset.   A spectral type of B2/3~II, with $M_V = -4.50$ and $(B-V)_0 = -0.22$,
would imply $E(B-V) = 0.35$ and a photometric distance of 563 pc.   In Table 2 we assume B3~IV and list 
$D_{\rm Shull} = 0.21$~kpc.  

\vspace{0.1cm}

\noindent
{\bf \#109 (HD 190429A).}  Our distance estimates, $D_{\rm Shull} = 2.57$~kpc and $D_{\rm GOS} = 2.38$~kpc, 
differ from previous values, 1.80~kpc and 2.9~kpc from Savage \etal\ (2001) and Bowen \etal\ (2008).   
We assume a SpT of O4~If (Sota \etal\ 2011), the same as Savage \etal\ (2001), and we adopt $M_V = -6.29$
for a Ib luminosity class.  The difference in distance appears to arise from the assumed photometry, $V$ and
$E(B-V)$.   The binary companion, HD~190429B (O9.5~II-III) is separated by $1.959''$ from HD~190429A.  We 
adopt  $B = 7.20$ and $V = 7.09$ for HD~190429A (Fabricius \etal\ 2002) with $E(B-V) = 0.43$, whereas 
Savage \etal\ (2001) adopt $V = 6.63$ presumably for the combined (A+B) system.   Our adopted $V = 7.09$ 
magnitude is consistent with the GOS photometry, which gives $V = 7.088$ for HD~190429A (Table~8 of 
Ma\'iz Apell\'aniz \etal\ 2004) and $V_J = 6.572$ with $A_V = 1.501$ for HD~190429AB (Ma\'iz Apell\'aniz \& 
Barb\'a 2018).  Bouret \etal\ (2012) analyze HD~190429A and derive a distance $2.45 \pm 0.20$~kpc based 
on $B = 7.201$, $V = 7.088$, $E(B-V) = 0.46$, and $M_V = -6.28$.  

\vspace{0.1cm}

\noindent
{\bf \#117 (HD 201345).}  Our distance estimates, $D_{\rm Shull} = 2.56$~kpc and $D_{\rm GOS} = 2.50$~kpc, 
are larger than the value, 1.91~kpc, in Savage \etal\ (2001).  We adopt a SpT of ON9.2~IV (Sota \etal\ 2014) 
with $M_V = -4.75$ compared to O9~V (Savage \etal\ 2001) with $M_V = -4.30$.   This produces a change
in absolute magnitude $\Delta M_V = -0.45$ (ON9.2~IV is more luminous).   



\clearpage




\startlongtable
\begin{deluxetable*}{rlrrrrcll}
\tablenum{1}
\tabletypesize{\scriptsize}
\tablecaption {Stellar Parameters and Photometry\tablenotemark{a} }
\tablecolumns{9}
\tablewidth{0pt}

\tablehead{ \colhead {ID} & \colhead {Target} & \colhead {$\ell$} & \colhead {$b$} & \colhead {$B$}
     & \colhead {$V$} & \colhead {$E(B-V)$} & \colhead {SpT} & \colhead {GOS}  \\
     &   & (deg) & (deg) & (mag) & (mag) & (mag) &  & ($V_J$, $A_{V_J}$)  }


\startdata
1    & {BD $35^{\circ}4258$}    & 77.19   & -4.74   &  9.42  & 9.41  & 0.31  &B0.5\,Vn      &                        \\
2    & {BD $53^{\circ}2820$}    &101.24  & -1.69   &10.05 & 9.95  & 0.40  & B0\,IVn        &                        \\
3    & {CPD-$59^{\circ}2600$} &287.60  & -0.74   & 8.82  & 8.61  & 0.53  & O6\,Vf          & (8.62, 2.19)    \\
4    & {CPD-$59^{\circ}2603$} &287.59  & -0.69   & 8.91  & 8.77  & 0.46  & O7\,Vnz       & (8.78, 1.61)    \\
5    & {CPD-$69^{\circ}1743$} &303.71  & -7.35   & 9.40  & 9.38  & 0.30  & B0.5\,IIIn     &     \\
6    & {CPD-$72^{\circ}1184$} &299.15  &-10.94  &10.61 &10.68 & 0.23 & B0\,III           &     \\
7    & {HD   3827}                       &120.79  &-23.23  & 7.76  &   8.01 & 0.02 & B0.7\,V        &    \\
8    & {HD   5005A}                     &123.12  & -6.24   & 8.19  & 8.10  & 0.41 & O4\,Vfc         & (8.10,1.44)   \\
\enddata
 
\tablenotetext{a} {Updated photometry and extinction ($V_J$, $A_{V_J}$) from the GOS survey
(Maiz Ap\'ell\'aniz \& Barb\'a 2018).   In multiple systems, the quoted magnitude is that of the 
brightest stellar component.   We analyze HD~93129A and HD~93206A as binary systems (Aa, Ab) --
see Gruner \etal\ (2019), Parkin \etal\ (2011), and further discussion in Appendix A. }

\end{deluxetable*}


\clearpage


\startlongtable
\begin{deluxetable*}  {rlllllccl}
\tabletypesize{\scriptsize}
\tablecolumns{9}
\tablecaption{Stellar Distance Estimates\tablenotemark{a} ($D_i$ in kpc) }
\tablenum{2}
\tablewidth{0pt}

\tablehead{ \colhead {ID} & \colhead {Star Name} & \colhead {SpT} & \colhead {Refs}
   & \colhead {$D_{\rm Gaia}$} & \colhead {$D_{\rm Bowen}$} & \colhead {$D_{\rm Savage}$}
   & \colhead {$D_{\rm Shull}$} & \colhead {$D_{\rm GOS}$}    \\
   &  &  &  & (DR2-2018) & (2008) & (2001) & (2019) & (2019) }
  
\startdata
 1    & {BD $35^{\circ}4258$}      & B0.5\,Vn       & 8,25     & 1.89[1.71,2.10]   &                          &               & 2.51 &             \\ 
 2    & {BD $53^{\circ}2820$}      & B0\,IVn         & 14,25   & 3.18[2.82,3.65]   & 4.8[3.6,6.4]     &               & 4.27 &             \\  
 3    & {CPD-$59^{\circ}2600$}   & O6\,Vf           & 2,14     & 5.11[4.28,6.36]   & 2.9[2.6,3.6]     & 2.84      & 2.71 &  2.11   \\
 4    & {CPD-$59^{\circ}2603$}   & O7\,Vnz       & 2,14      & 3.66[3.22,4.22]   & 3.5[3.1,4.4]     & 2.68      & 2.81 &  2.60   \\
 5    & {CPD-$69^{\circ}1743$}   & B0.5\,IIIn      & 10,11   & 3.43[3.04,3.93]   &                          & 4.68      & 4.47 &              \\                               
 6    & {CPD-$72^{\circ}1184$}   & B0\,III            & 16         & 6.88[5.46,9.32]  &                          & 9.41      & 9.85 &              \\  
 7    & {HD   3827}                         & B0.7\,V         & 26         & 2.15[1.85,2.59]   &                          & 2.04      & 1.88 &              \\  
 8    & {HD   5005A}                       & O4\,Vfc         & 1,8       & 2.71[2.39,3.14]   & 2.9[2.8,3.1]     & 2.28      & 3.13 &  2.89    \\
\enddata
\end{deluxetable*}


\clearpage


\startlongtable
\begin{deluxetable*}{rlllrllccc}
\tablecolumns{10}
\tabletypesize{\scriptsize}
\tablenum{3}
\tablewidth{0pt}
\tablecaption{Parallax Errors and Distance Estimates\tablenotemark{a} ($D_i$ in kpc) }

\tablehead{ \colhead {ID} & \colhead {Star Name} & \colhead {SpT }& \colhead {Parallax} & \colhead {Ratio}
   & \colhead {$G_{\rm Gaia}$} & \colhead {$D_{\rm Gaia}$ }& \colhead{$D_{\rm Shull}$ } 
   &\colhead {$D_{\rm GOS}$ } &\colhead {Ratio}   \\
   \colhead{}
& \colhead{}
& \colhead{} 
& \colhead{$\varpi \pm \sigma_{\varpi}$(mas)}
& \colhead{$\varpi / \sigma_{\varpi}$}
&\colhead{(mag)} 
& \colhead{(DR2-2018)} 
& \colhead{(2019)}
& \colhead{(2019)}
& \colhead {Gaia/GOS}
}
     
\startdata
   3   & {CPD-$59^{\circ}2600$}    & O6\,Vf          & $0.1655\pm0.0382$ & 4.3 & 8.46 & 5.11[4.28-6.36] & 2.71 &  2.11 & 2.42  \\
   4    & {CPD-$59^{\circ}2603$}   & O7\,Vnz       & $0.2435\pm0.0368$ & 6.6 & 8.69 & 3.66[3.22-4.22] & 2.81 &  2.60 & 1.41   \\
   8    & {HD   5005A}                       & O4\,Vfc        & $0.3388\pm0.0500$ & 6.8 & 8.52 & 2.71[2.39-3.14] & 3.13 &  2.89 & 0.94  \\
   9    & {HD  12323}                        & ON9.2\,V     & $0.3557\pm0.0442$ & 8.1 & 8.90 & 2.59[2.33-2.93] & 2.81 &  3.04 & 0.85   \\ 
 10   & {HD  13268}                        & ON8.5\,IIIn  & $0.5906\pm0.0466$ &12.7 & 8.11 & 1.61[1.50-1.74  & 2.77 &  2.80 & 0.58  \\
 11   & {HD  13745}                        & O9.7\,IIn      & $0.4405\pm0.0528$ &  8.3 & 7.78 & 2.13[1.91-2.39] & 2.80 &  2.81 & 0.76  \\
 12   & {HD  14434}                        & O5.5\,Vnfp  & $0.3912\pm0.0473$ &  8.3 & 8.43 & 2.37[2.13-2.67] & 2.95 &  2.98 & 0.85 \\
 13   & {HD  15137}                        & O9.5\,II-IIIn  & $0.2700\pm0.0502$ & 5,4  & 7.81 & 3.33[2.86-4.00] & 2.88 &  3.00 & 1.11  \\
\enddata
      
\tablenotetext{a} {Of  the 84 stars with GOS classification and photometry, 81 have reliable {\it Gaia}-DR2 parallaxes,
 quantified by $\varpi / \sigma_{\varpi}$, the inverse of the relative parallax errors (columns 4 and 5). Column~6 lists $G-band$
 magnitudes from {\it Gaia}-DR2, which range from $G = 4.94$ (\#127) to $G = 9.71$ (\#62) with $\langle G \rangle =  7.51$. 
 Columns 7, 8, 9 list distances (kpc) from {\it Gaia} and two photometric values, $D_{\rm Shull}$ and $D_{\rm GOS}$,
 from Table 2.   Column~10 lists the parallax-to-photometric distance ratio $(D_{\it Gaia} / D_{\rm GOS})$, plotted in 
 Figure~5 vs.\ a quality indicator ($\varpi / \sigma_{\varpi}$) of the parallax measurement. } 
      
\end{deluxetable*}


\clearpage


\begin{deluxetable*}{llllcccccll}
\tablecolumns{11}
\tabletypesize{\scriptsize}
\tablenum{4}
\tablewidth{0pt}
\tablecaption{Carina OB1 Distances\tablenotemark{a}  }

\tablehead{ \colhead {Star (ID)} & \colhead {SpT } & \colhead {$V_{J,0}$} & \colhead {$A_{V_J}$ }& \colhead {$M_V^{(B)}$ }
   & \colhead {$M_V^{(M)}$ }& \colhead {$D_{\rm Gaia}$ } & \colhead{$D_{\rm phot}$ }& \colhead{$D_{\rm GOS}$ } 
   & \colhead {$D'_{\rm GOS}$ } & \colhead {Cluster} \\
   &  & (mag) & (mag) & (mag) & (mag) & (kpc) & (kpc) & (kpc) & (kpc) &   }

\startdata
 {HD 93204 (\#44)}                     & O5.5\,V            & 6.827  & 1.611  & $-5.35$ & $-5.07$ & 2.09[1.94,2.27]  & 2.74  & 2.73    & 2.39  & Tr~16  \\ 
 {HD 93205 (\#45)}                     & O3.5\,V            & 6.203  & 1.539  & $-5.73$ & $-5.65$ & 2.49[2.27,2.76]  & 2.38  & 2.43    & 2.34  & Tr~16  \\
 {HD 93250 (\#48)}                     & O4\,III               & 5.515  & 1.850  & $-6.20$ & $-5.98$ & 2.54[2.36,2.75]  & 2.14  & 2.20    & 1.99  & Tr~16  \\
 {HD 303308 (\#137)}                & O4.5\,V             & 6.547  & 1.572  & $-5.57$ & $-5.36$ & 2.29[2.10,2.51]  & 2.43  & 2.66    & 2.40  & Tr~16  \\
 {CPD-$59^{\circ}2600$ (\#3)} & O6\,V                & 6.423  & 2.193  & $-5.20$ & $-4.92$ & 5.11[4.28,6.36]  & 2.18  & 2.11    & 1.86  & Tr~16  \\        
 {CPD-$59^{\circ}2603$ (\#4)} & O7\,V                & 7.171  & 1.607  & $-4.90$ & $-4.63$ & 3.66[3.22,4.22]  & 2.32  & 2.60    & 2.29  & Tr~16  \\
 {CPD-$59^{\circ}2591$}          & O8.5\,V             & 7.682  & 3.123  & $-4.45$ & $-4.19$ & 2.32[2.10,2.58]  & 2.83  & 2.67    & 2.37  & Tr~16  \\
 {CPD-$59^{\circ}2626$}          & O7.5\,V             & 6.999  & 2.993  & $-4.75$ & $-4.48$ & 3.84[3.40,4.41]  & 2.40  & 2.24    &1.98   & Tr~16  \\
\enddata

\tablenotetext{a} {Photometric distances and offset-corrected {\it Gaia} parallax distances for 29 stars 
associated with clusters (Trumpler~16, Trumpler~14, Trumpler~15, Collinder~228) in the Carina Nebula
(Humphreys 1978; Alexander \etal\ 2016).  We list $D_{\rm Gaia}$ (column 7) and three photometric 
distance estimates denoted:  $D_{\rm phot}$ (column 8), $D_{\rm GOS}$ (column 9), and $D'_{\rm GOS}$ 
(column 10).  $D_{\rm phot}$ is calculated using $V$ and $E(B-V)$, as in Table 2, but with anomalous 
extinction, $R_V = A_V/E(B-V)= 4.0$ rather than $R_V = 3.1$.  $D_{\rm GOS}$ is based on GOS digital
photometry and extinction and the absolute magnitude grid ($M_V$) from Bowen \etal\ (2008).   
$D'_{\rm GOS}$ uses absolute magnitudes from Martins \etal\ (2005).  Column~2 gives updated SpTs 
from the GOS spectroscopic survey (Sota \etal\ 2011, 2014).  Columns 3 and 4 list the extinction-corrected 
visual magnitude ($V_{J,0} = V_J - A_{V_{J}}$) and visual extinction $A_V$ from 
Ma\'iz Apell\'aniz \& Barb\'a (2018).   
Column 5 lists absolute magnitudes ($M_V$) from Bowen \etal\ (2008) used for $D_{\rm phot}$ and
$D_{\rm GOS}$.   Although the multiple system HD~93160 has previously been associated with 
Tr~16 (Walborn 1973b; Humphreys 1978), it more likely belongs in Tr~14  (see Fig.\ 12 of Sota \etal\ 2014). 
HD~93129A (ID \#42) is a binary system (Gruner \etal\ 2019) whose GOS photometry is combined with 
HD~93129B; we exclude it from the statistics.  For individual clusters, we find mean distances 
$\langle D_{\rm GOS} \rangle$ for Tr~16 ($2.55 \pm 0.30$~kpc), Tr~14 ($2.68 \pm 0.31$~kpc), and
Coll~228 ($2.58 \pm 0.23$~kpc).   Mean distances for the ensemble of all 29 O-stars are listed at the 
bottom of the table, with consistent photometric distances 
$\langle D_{\rm phot} \rangle = 2.62 \pm 0.29$~kpc and 
$\langle D_{\rm GOS} \rangle = 2.60 \pm 0.28$~kpc.
}
 
\end{deluxetable*}



\begin{deluxetable*}{llcccccccc}
\tablecolumns{10}
\tabletypesize{\scriptsize}
\tablenum{5}
\tablewidth{0pt}
\tablecaption{Perseus OB1 Distances\tablenotemark{a}  }

\tablehead{ \colhead {Star (ID)} & \colhead {SpT } & \colhead {$V_{J,0}$} & \colhead {$A_{V_J}$ }
   & \colhead {$M_V^{(B)}$ } & \colhead {$M_V^{(M)}$ } & \colhead {$D_{\rm Gaia}$ }
   & \colhead {$D_{\rm phot}$ } & \colhead {$D_{\rm GOS}$ } & \colhead {$D'_{\rm GOS}$ }  \\
   &  & (mag) & (mag) & (mag) & (mag) & (kpc) & (kpc) & (kpc) & (kpc)  } 

\startdata
 {HD 12323 (\#9)}    &  ON9.2\,V    & 8.176  & 0.736  & $-4.24$  & $-3.99$ & 2.59[2.33,2.93]  & 2.81   & 3.04  & 2.71   \\
 {HD 13268 (\#10)}  &  ON8.5\,III    & 6.836  & 1.340  & $-5.40$ & $-5.32$ & 1.61[1.50,1.74]   & 2.77  & 2.80 & 2.70   \\
 {HD 13745 (\#11)}  &  O9.7\,II        & 6.415  & 1.449  & $-5.83$ & \dots       & 2.13[1.91,2.39]  & 2.80   & 2.81 & \dots   \\
 {HD 14434 (\#12)}  &  O5.5\,V       & 7.023  & 1.467  & $-5.35$ & $-5.07$  &  2.37[2.13,2.67]  & 2.95  & 2.98 & 2.62   \\
 {HD 15137 (\#13)}  &  O9.5\,II-III   & 6.865  & 1.007  & $-5.52$ & \dots        & 3.33[2.86,4.00]  & 2.88   & 3.00 & \dots  \\
 {HD 15642 (\#15)}  &  O9.5\,II-III   & 7.432   & 1.098 & $-5.52$ & \dots        & 3.69[3.20,4.34]  & 3.75   & 3.89 & \dots   \\
 {HD 13022}              &  O9.7\,II-III   & 6.946   & 1.835 & $-5.47$ & \dots        & 2.26[2.04,2.52]  & 3.35   & 3.04 & \dots   \\      
 {HD 12993}              &  O6.5\,V       & 7.512   & 1.454 & $-5.05$ & $-4.77$  & 2.41[2.17,2.70]  & 3.27   & 3.25 & 2.86    \\
\enddata 

\tablenotetext{a} {Same information as in Table 3 for 12 stars in Per~OB1.  As in Table 2, photometric distances 
$D_{\rm phot}$ were calculated from ($B$, $V$) photometry, SpTs ($M_V$), and extinction ($A_V$) assuming the 
standard value $R_V = A_V/E(B-V) = 3.1$.  We present two GOS-based photometric distances, $D_{\rm GOS}$ and 
$D'_{\rm GOS}$, with absolute magnitudes $M_V^{(B)}$ Bowen \etal\ (2008) and $M_V^{(M)}$ (Martins \etal\ 2005) 
respectively.  In four cases (ellipses) SpTs are off the Martins \etal\ grid.  Mean photometric distances based on the
Bowen \etal\ (2008) grid are consistent with $\langle D_{\rm phot} \rangle = 2.95 \pm 0.23$~kpc and
$\langle D_{\rm GOS} \rangle = 2.99 \pm 0.14$~kpc.  The parallax distance is more uncertain, 
$\langle D_{\rm Gaia} \rangle = 2.47 \pm 0.57$~kpc. 
For these statistics, we exclude three stars unlikely to be association members (Lee \& Lim 2008):
HD~15642 and HD 14442 (larger distances) and HD~14633, located at a much different Galactic latitude 
($b = -18.20^{\circ}$).   } 

\end{deluxetable*}

\end{document}